\documentclass[12pt,epsfig,aps]{article}
\usepackage{graphicx}
\usepackage{epsfig}
\usepackage{color}
\usepackage{amssymb,latexsym,amsmath}

\newcommand{\bs}{\mathbf}

\begin{document}
\title{ \Large{ \bf Hadronic Reaction Zones in Relativistic Nucleus-Nucleus Collisions} }
\author{
D. Anchishkin$^1$, V. Vovchenko$^2$, S. Yezhov$^2$
\\ $^1${\small \it Bogolyubov Institute for Theoretical Physics,
             03680 Kiev, Ukraine}
\\ $^2${\small \it Taras Shevchenko Kiev National University, 03022 Kiev,
Ukraine} }

\maketitle

\begin{abstract}
On the basis of the proposed algorithm for calculation of the hadron reaction
rates, the space-time structure of the relativistic nucleus-nucleus collisions is
studied.
The reaction zones and the reaction frequencies for various types of reactions are
calculated for AGS and SPS energies within the microscopic transport model.
The relation of the reaction zones to the kinetic and chemical freeze-out processes
is discussed.
It is shown that the space-time freeze-out layer is most extended
in time in the central region, while, especially for higher collision energies,
the layer becomes very narrow at the sides.
The parametrization of freeze-out hypersurface
in the form of specific hyperbola of constant proper time was confirmed.
The specific characteristic time moments of the fireball evolution are introduced.
It is found that the time of the division of a reaction zone into two separate parts
does not depend on the collision energy.
Calculations of the hadronic reaction frequency show that the evolution of nucleus-nucleus collision can be divided into
two hadronic stages.
\end{abstract}

%%%%%%%%%%%%%%%%%%%%%%
\section{Introduction}
\label{intro}
%%%%%%%%%%%%%%%%%%%%%%
In the heavy-ion collisions, a strongly interacting hadronic system is formed --
a fireball.
This system is identified with a space-time region, in which the intensive
hadronic reactions are running.
With regard for a model describing the system, we can separate such stages of
the fireball evolution as the formation of a fireball, its
thermalization, hydrodynamic expansion, freeze-out, etc.
The detailed knowledge of space-time structure of
a fireball
can serve as a way for formulation and selection of
models used to address different problems in the field.

As parameters for the determination of the stages of evolution of a system, one
can take the particle density $n(t,\bs r)$ \cite{CERES,Anchishkin:2012jk},
energy density $\epsilon(t,\bs r)$ \cite{Anchishkin:2012jk,RusskikhIvanov2006},
temperature $T(t,\bs r)$ \cite{mclerran1986,Huovinen2007},
mean free path, rate of collisions of particles $\Gamma(t,\bs r)$, etc.
The same parameters are also used for the determination of the chemical and
kinetic freeze-out processes \cite{Cleymans2006,Abelev2009,Aggarwal2011}.
Depending on the chosen parameter, one can get various fireball representations.
In the present work, we use the hadron reaction rate $\Gamma(t,\bs r)$
(number of reactions in a unit volume per unit time)
in a given four-dimensional region of the
space-time as a parameter of the spatial evolution of the interacting system.
Such a quantitative estimate allows one to define the reaction
zone \cite{Anchishkin2010}, which study gives a
possibility to establish the space-time structure of a fireball from
the viewpoint of the interaction (collision) intensity at every point of the
space-time.
Then, the regions of a fireball can be distinguished by the interaction
intensity that can be characterized by the number of collisions in a
unit volume of the space-time.
We use this quantity to estimate and to quantify various regions of the reaction
zone.

Another important question which can be clarified by the study of the zone of
reactions is how the space-time boundary of a particular region of the reaction
zone is related to the processes of kinetic and chemical freeze-outs.
Since the kinetic freeze-out is the process of establishment of a final
distribution of hadrons in the momentum space, the sharp kinetic
freeze-out hypersurface is an imaginary hypersurface, outside of which
there are no collisions between radiated hadrons (or a very small amount of
collisions is admitted).
This kind of picture is an idealized description for heavy-ion collisions,
and studies within microscopic models have shown
that freeze-out happens in the extended space-time domain rather
than on some 3-dimensional space-time hypersurface \cite{Bravina1995,Bravina1999,Bass1999},
however, the sharp freeze-out hypersurface is frequently used
within the Cooper-Frye prescription \cite{cooper-frye-PRD-v10-1974},
e.g. for calculations
of the hadron spectra in hydrodynamical description or for making transition from
the fluid dynamical stage of collision to the stage of dilute hadron gas in hybrid models.
In this sense, the space-time boundary of a reaction zone and the sharp
kinetic freeze-out hypersurface can be put in correspondence.
%%%%%%%%%%%%%%%%%%%%%%%%%%%%%%%%%%%%%%%%%%%%%%%%%%%%%%%%%%%%%%%%%%%%%%%%%%%%%%%%

%%%%%%%%%%%%%%%%%%%%%%%%
\section{Reaction zones}
\label{sec:1}
%%%%%%%%%%%%%%%%%%%%%%%%
The number of reactions $N_{\rm coll}$ in the given space-time region $\Omega$
can be evaluated in the following way:
\begin{equation}\label{Ncoll}
N_{\rm coll}(\Omega) = \int_\Omega d^4x \, \Gamma(x) \, ,
\end{equation}
where the four-density of reactions $\Gamma(x)$ can be evaluated, in turn, in the
framework of a certain model approximation, e.g., like that in
\cite{TomasikWiedeman2003,HungShuryak1998}.
In particular, $\Gamma(x)$ can be calculated with the use of a distribution
function $f(x,p)$ within a transport model.

The reaction zone is defined as {\it the space-time region where a certain
fraction of all reactions of a certain type take
place} \cite{Anchishkin2010}.
This space-time region is chosen so that it should be the most intense with
respect to the reaction rate, i.e., it has the smallest possible volume.
Reactions are classified by the type and the number of particles taking part in
these reactions (see Table \ref{tab:reacttype}).
The reaction zone can be calculated for various types of reactions with the use
of the particular rate for a given reaction type.

One can get a feeling of the definition given above, while associating the reaction
zone with the flame of a candle (see Fig.~\ref{fig:Zor-trz}).
Indeed, it is the region that is the most intense with respect to exothermic
``luminous'' reactions and is basically a reaction zone (as usual, we do not care
about the regions which contain the products of the oxidizing reaction that are
not visible to us).

In the present paper, we investigate the most central nucleus-nucleus collisions.
Because of the symmetry of central collisions, the reaction density does not
depend on the azimuthal angle $\varphi$ in the $x$-$y$ plane in the
cylindrical coordinates, i.e., $\Gamma(t,x,y,z) = \Gamma(t,r,z)$.
In this case it is possible to build a three-dimensional reaction zone in
coordinates $(t,r,z)$, where $r=\sqrt{x^2+y^2}$.

Our task is to determine a hypersurface that confines the
volume containing a certain part $\alpha \ \ (0<\alpha<1)$ of the total
number of all hadronic reactions $N_{\rm tot}$.
This hypersurface can be determined by the equation
\begin{equation}
\Gamma(t,\,r,\,z)\ =\ \Gamma_c \,,
\label{eq:gammac-alpha}
\end{equation}
where the critical value $\Gamma_c$ satisfies the equation for a given value
of $\alpha$
\begin{equation}
\int\limits dt \, dx \, dy \, dz \, \Gamma(t,x,y,z) \operatorname{\theta}(\Gamma-\Gamma_c)
= 2\pi \int\limits dt \, dr \, dz \, r \, \Gamma(t,r,z) \operatorname{\theta}(\Gamma-\Gamma_c)
=\alpha N_{\rm tot}\,.
\label{eq:gammac-alpha}
\end{equation}
It is obvious that the value of $\Gamma_c$ ``determines'' the ``luminous''
part of the reaction zone analogously to the luminous
region of the flame of a candle.

%%%%%%%%%%%%%%%%%%%%%%%%%%%%%%%%%%%%%%%%%%%%%%%%%%%%%%%%%%%%%%%%%%%%%%%%%%%%%%%%
\begin{table}
 \caption{{\bf Classification of reactions}}      %title
 \centering                                                             %centering table
 \begin{tabular}[c]{l|l|l}                                              %columns {format}
 \hline\hline                                                           %horizontal lines
 1 & $1 \to 2' + m, \,\, m \geq 0$ & decay \\
 \hline
 2 & $2 \to 1'$                    & fusion \\
 \hline
 3 & $2 \to 2$                     & elastic scattering \\
 \hline
 4 & $2 \to 2' + m, \,\, m \geq 0$ & inelastic reaction \\
 \hline
 \end{tabular}
 \label{tab:reacttype}
\end{table}
%%%%%%%%%%%%%%%%%%%%%%%%%%%%%%%%%%%%%%%%%%%%%%%%%%%%%%%%%%%%%%%%%%%%%%%%%%%%%%%%

To carry out evaluations, we use the microscopic transport model UrQMD v2.3
\cite{UrQMD1998,UrQMD1999}, which allows one to calculate the four-density of
reactions at every point of the space-time and to select the reactions of a necessary
type and the species of particles.
We take a large four-volume of reactions $C_{R}$, which is determined in the cylindrical
coordinates $(t,r,z)$ as: $0<t<200~{\rm fm}/c$, $0<r<200~{\rm fm}$
and $-200~{\rm fm}<z<200~{\rm fm}$.
%
%%%%%%%%%%%%%%%%%%%%%%%%%%%%%%%%%%%%%%%%%%%%%%%%%%%%%%%%%%%%%%%%%%%%%%%%%%%%%%%%
\begin{figure}
\begin{center}
\includegraphics[width=0.7\textwidth] {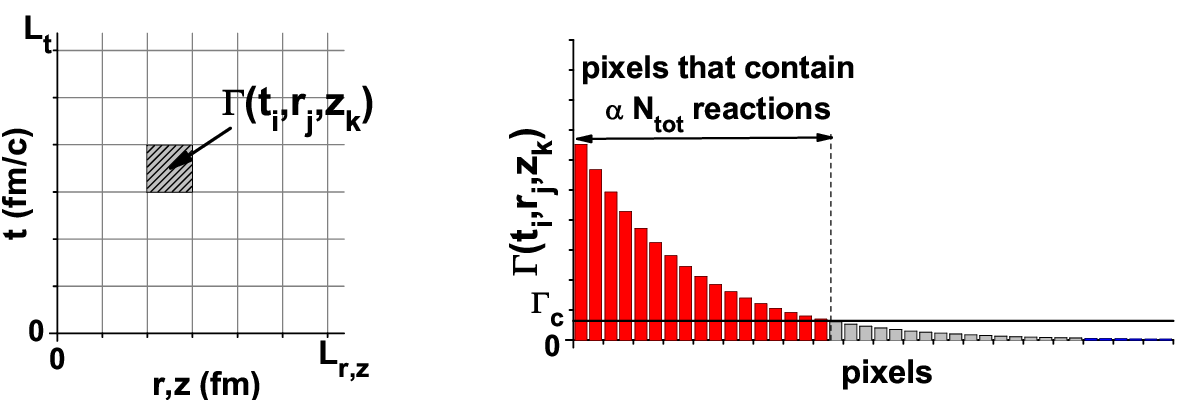}
\caption{Algorithm for the selection of pixels in the determination of
the reaction zone.
Left panel: the grid of pixels in the cylindrical coordinates $(t,r,z)$.
Right panel: the arrangement of all bins, which are put in correspondence to
pixels, in a linear hierarchy in accordance with their height (the density of
reactions $\Gamma$ in the corresponding pixel).
A value of $\Gamma_c$ defines the lower (right) boundary, when we go from the
very left bin (pixel with the highest density $\Gamma$) to the right along this
hierarchy, see Eq.~\eqref{eq:gammac-alpha}. }
\label{fig:ZoR-method}
\end{center}
\end{figure}
%%%%%%%%%%%%%%%%%%%%%%%%%%%%%%%%%%%%%%%%%%%%%%%%%%%%%%%%%%%%%%%%%%%%%%%%%%%%%%%%

To calculate the critical rate $\Gamma_c$ and to determine the reaction zone, we
divide the four-volume $C_R$ into cells (pixels), i.e., elements of the
four-space, as shown in Fig.~\ref{fig:ZoR-method} (left panel).
Let $\Omega_{ijk}=\Omega(t_i, r_j, z_k)$ be the four-volume of a pixel with
coordinates $(t_i, r_j, z_k)$ of its center.
Hence, this pixel is defined as
\begin{equation}
t_i-\frac12 \Delta t < t < t_i+\frac12 \Delta t \,, \quad
r_j-\frac12 \Delta r < r < r_j+\frac12 \Delta r\,, \quad
z_k-\frac12 \Delta z < z < z_k+\frac12 \Delta z\,,
\label{eq:pixel}
\end{equation}
where $\Delta t = \Delta r = \Delta z = 1$~fm in our calculations.
The values of $\Delta$ determine the sampling accuracy in our method.
The four-volume of a pixel with coordinates $(t_i, r_j, z_k)$
is $\Omega(t_i, r_j, z_k) = 2\pi r_j \, \Delta t \Delta r \Delta z$.
Then, for each three numbers $(t_i, r_j, z_k)$, which determine a certain
pixel, we can calculate the reaction density $\Gamma(t_i, r_k, z_k)$ in the
given pixel from UrQMD by calculating the number of reactions in that pixel
and dividing it by the four-volume $\Omega(t_i, r_j, z_k)$ of the pixel.
After this, we put all pixels on the line in the form of bins of equal width
and of a height, which is equal to the pixel intensity $\Gamma(t_i, r_k, z_k)$.
We sort the bins (pixels) from left to right by the following hierarchy:
from a pair of bins (pixels), the left bin is higher (the pixel has a larger
reaction density), see Fig.~\ref{fig:ZoR-method} (right panel).
The total integral of $\Gamma(t, r, z)$ (the sum of $\Gamma$-values in pixels
multiplied by corresponding pixel four-volume $\Omega$) is equal to the total
number of all hadron reactions $N_{\rm tot}$ in the four-volume of reactions
$C_R$.

Let us sum the values of reaction densities multiplied by the corresponding
$\Omega$ beginning from the left according to the obtained hierarchy.
We recall that the value of each $\Gamma(t_i,r_j,z_k)\,\Omega(t_i,r_j,z_k)$
gives the number of reactions in the corresponding pixel.
Increasing the number of pixels in the sum, we can reach the value of sum
that is equal to a given number $\alpha\, N_{\rm tot}$, where $\alpha$ is a
given fraction of the absolute number of all reactions $N_{\rm tot}$, see
Fig.~\ref{fig:ZoR-method} (right panel) and Eq.~\eqref{eq:gammac-alpha}.
For instance, for $\alpha=0.9$ the critical density of reactions $\Gamma_c$ will
corresponds to the $90$\% of the total number of reactions.
The current value of reaction density in the hierarchy at that point is
$\Gamma_c (\alpha)$.

%%%%%%%%%%%%%%%%%%%%%%%%%%%%%%%%%%%%%%%%%%%%%%%%%%%%%%%%%%%%%%%%%%%%%%%%%%%%%%%%
\begin{figure}
\centering
  \includegraphics[width=.15\textwidth]{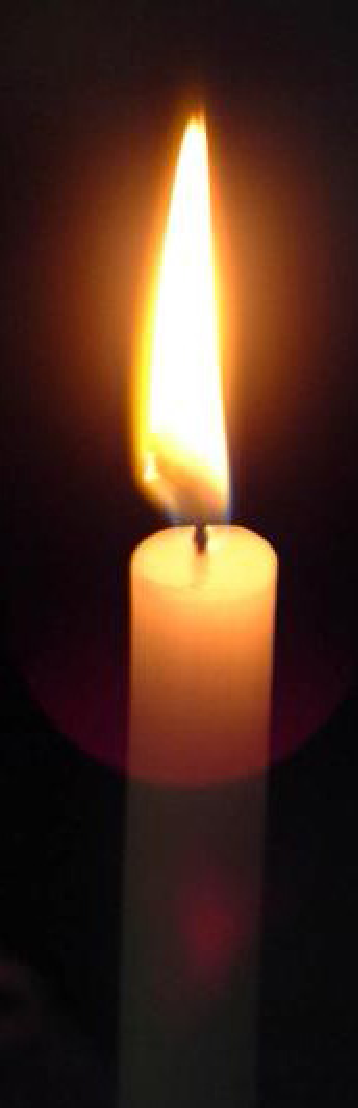}
  \qquad \quad
  \includegraphics[width=.65\textwidth]{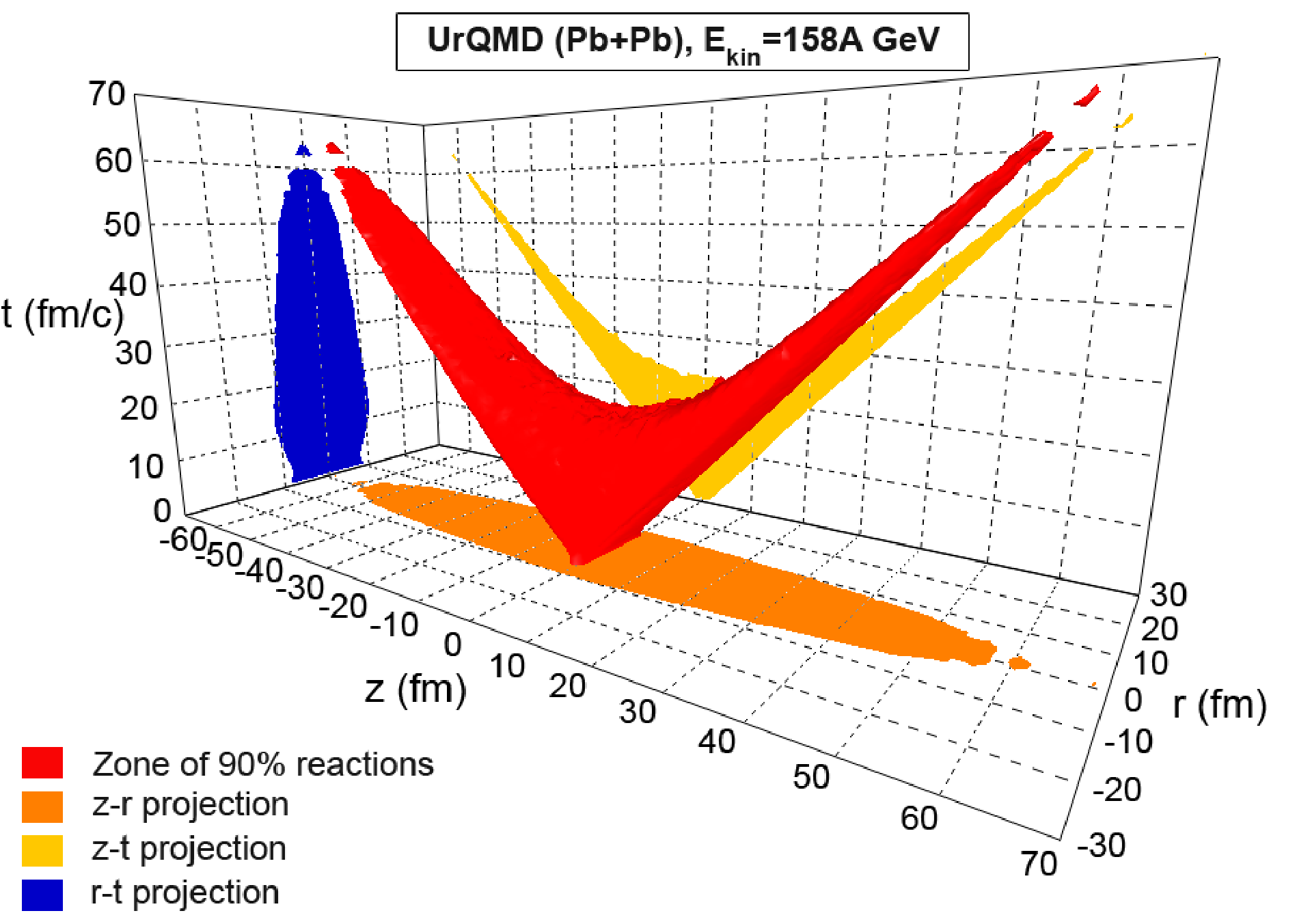}\\
  \caption{Left -- the flame of a candle as an analogue of the region, where reactions take place.
Right -- the three-dimensional reaction zone that determines the space-time
region, where 90\% of all hadronic reactions
under SPS conditions ($E_{\rm kin} = 158$A GeV) take place.}
\label{fig:Zor-trz}
\end{figure}
%%%%%%%%%%%%%%%%%%%%%%%%%%%%%%%%%%%%%%%%%%%%%%%%%%%%%%%%%%%%%%%%%%%%%%%%%%%%%%%%
In Fig. \ref{fig:Zor-trz}, we show the calculation results under conditions at the CERN
Super Proton Synchrotron (SPS), Pb+Pb at 158A GeV in the case of central
collisions.
The figure depicts the space-time region, where 90\% of all hadronic reactions take
place.
For this calculations, we use the coordinates ($t,r,z$), where $r=\pm \sqrt{x^2+y^2}$.
In addition, we show the different orthographic projections of the reaction zone on the
coordinate planes $z$-$t$ (yellow), $r$-$t$ (blue), and $z$-$r$ (orange).

\begin{figure}[!t]
\centering
\begin{minipage}{.48\textwidth}
\includegraphics[width=\textwidth]{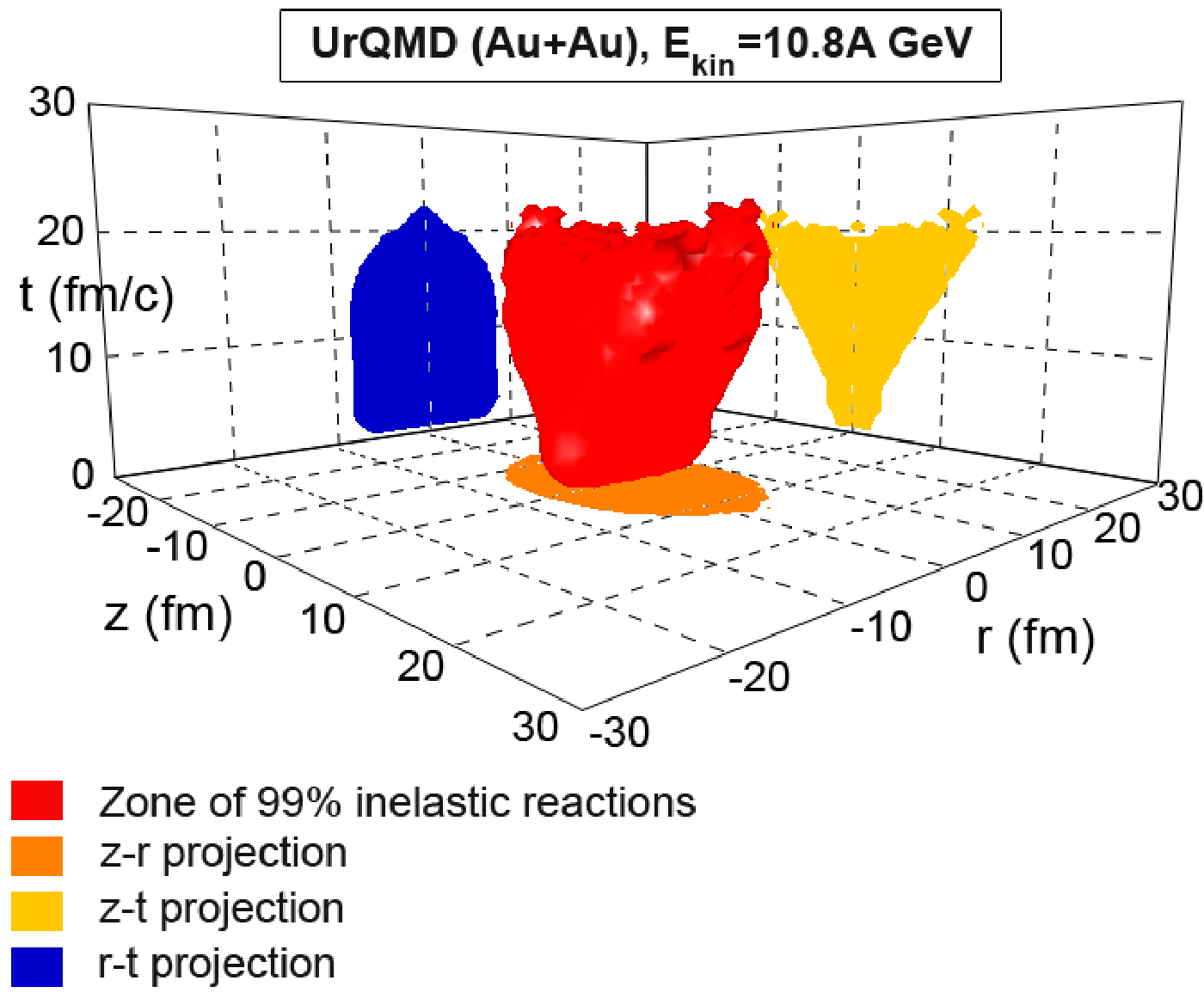}
\end{minipage}
\begin{minipage}{.48\textwidth}
\includegraphics[width=\textwidth]{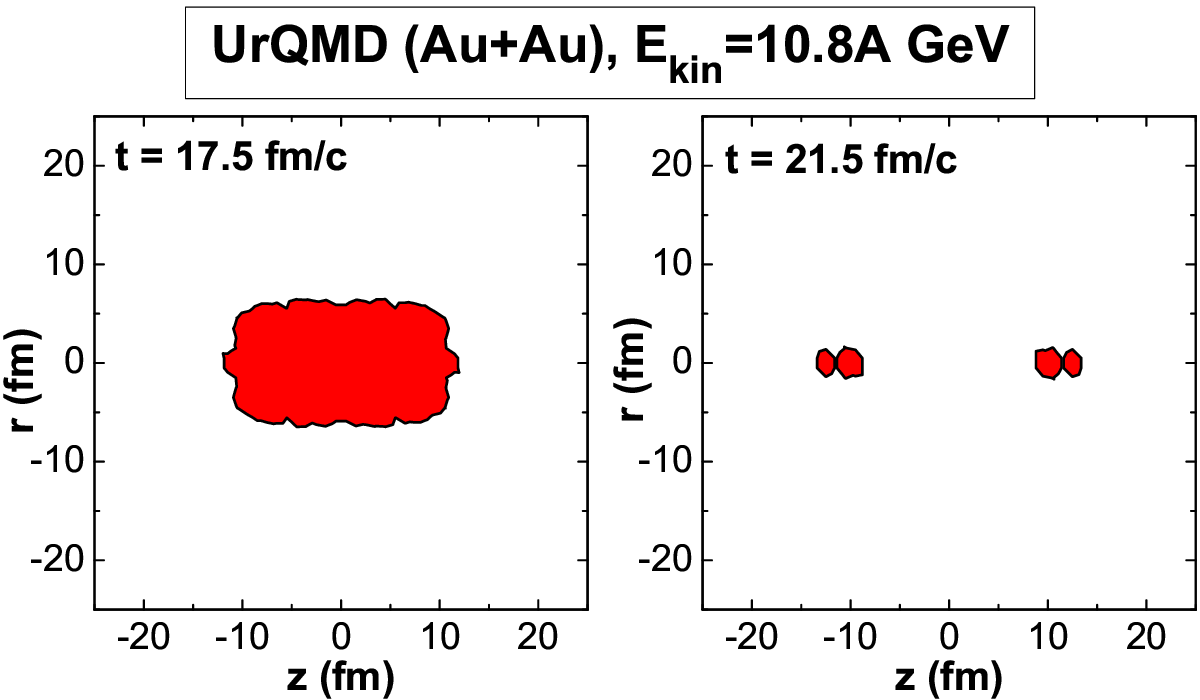}
\end{minipage}
\caption{Left -- the three-dimensional reaction zone, which determines the
space-time region where 99\% of all inelastic hadronic reactions under AGS
conditions ($E_{\rm kin} = 10.8$A GeV) take place.
Right -- the same reaction zone in the $z$-$r$ coordinates at different times before
and after the hot-fireball break-up. }
\label{fig:ZoR-3D-AGS}
\end{figure}

\begin{figure}[!t]
\centering
\begin{minipage}{.48\textwidth}
\includegraphics[width=\textwidth]{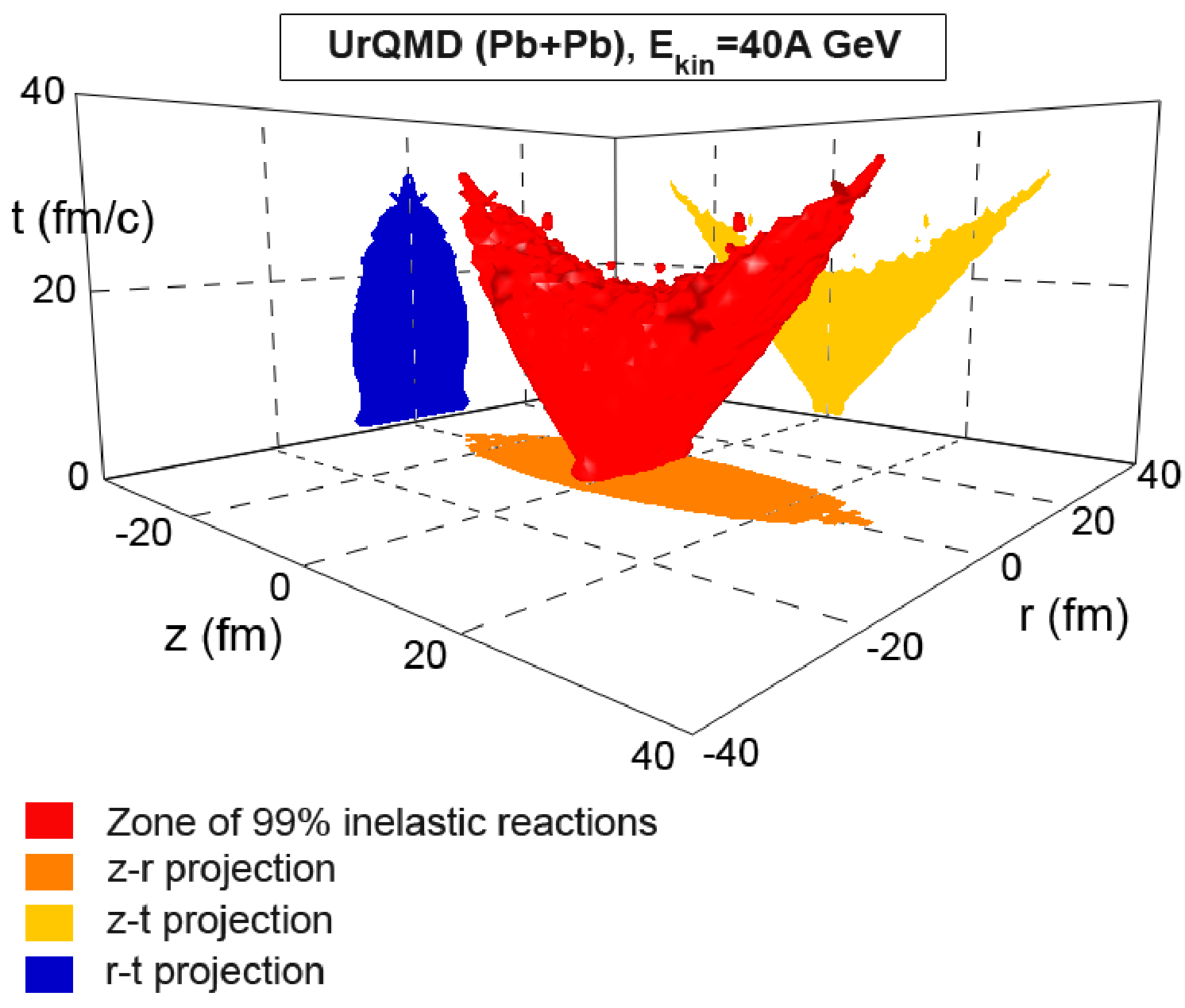}
\end{minipage}
\begin{minipage}{.48\textwidth}
\includegraphics[width=\textwidth]{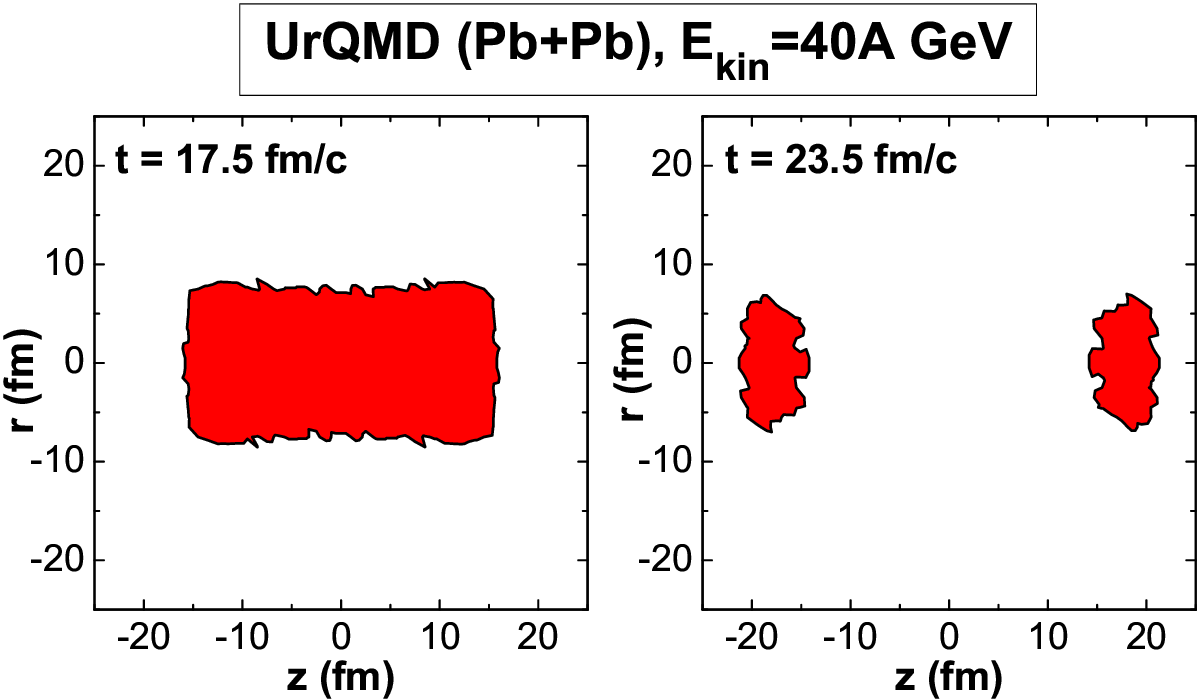}
\end{minipage}
\caption{Same as in Fig. \ref{fig:ZoR-3D-AGS}, but
under SPS conditions at $E_{\rm kin} = 40$A GeV.}
\label{fig:ZoR-3D-lowSPS}
\end{figure}

\begin{figure}
\centering
\begin{minipage}{.48\textwidth}
\includegraphics[width=\textwidth]{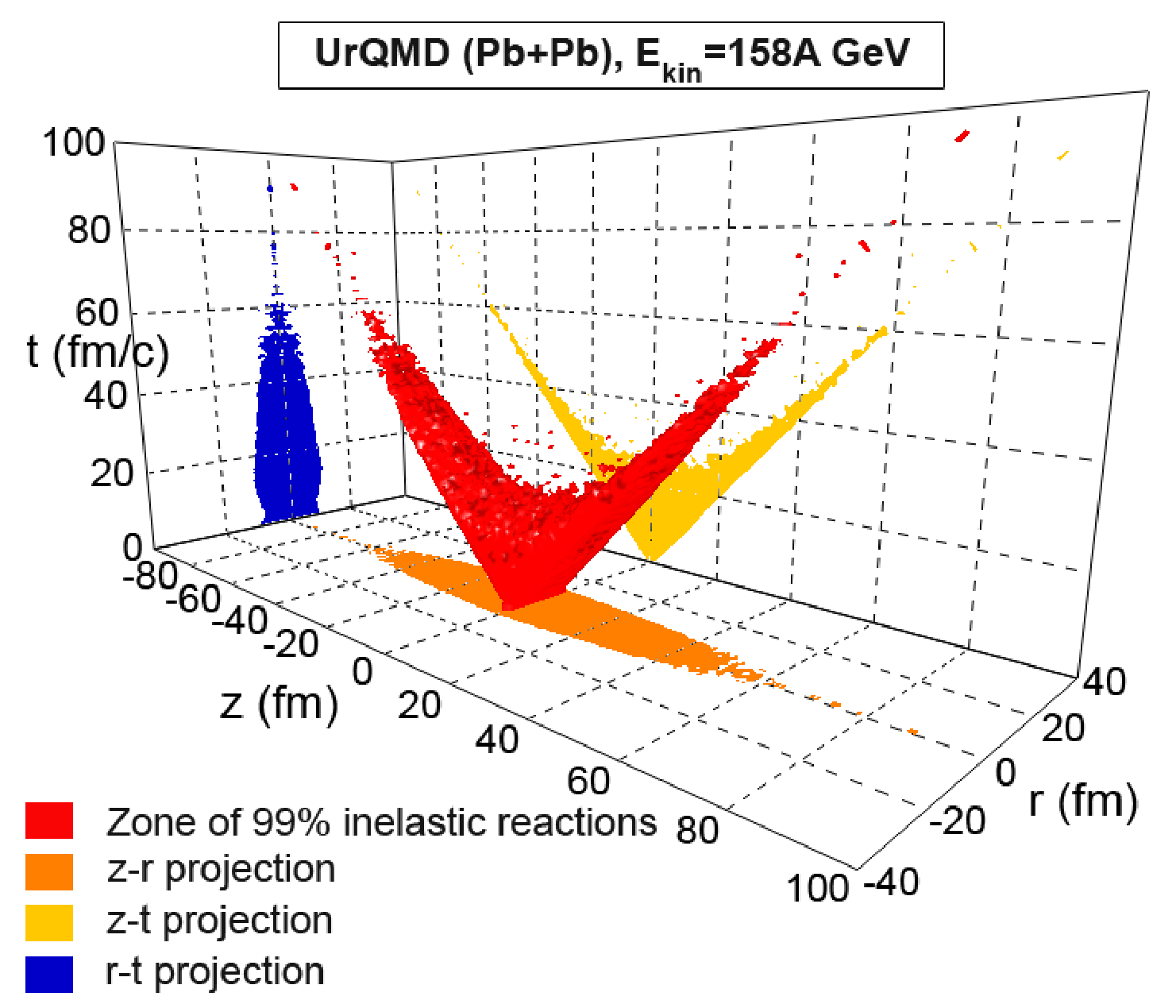}
\end{minipage}
\begin{minipage}{.48\textwidth}
\includegraphics[width=\textwidth]{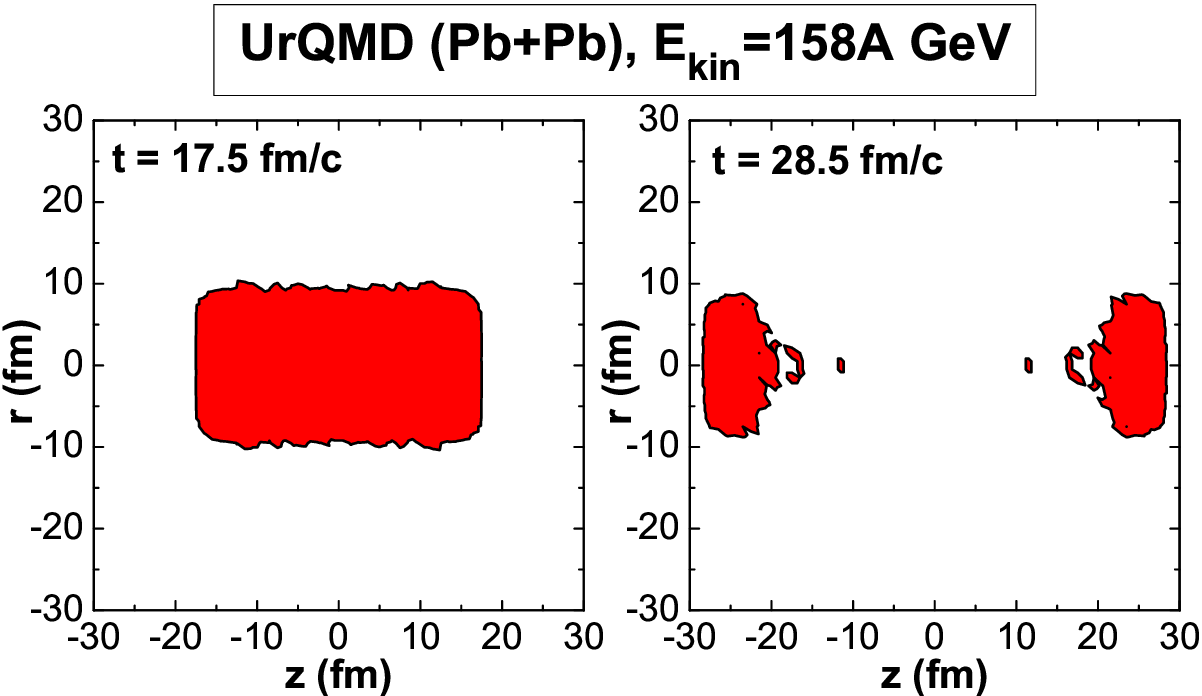}
\end{minipage}
\caption{Same as in Fig. \ref{fig:ZoR-3D-AGS} but
for SPS conditions at $E_{\rm kin} = 158$A GeV.}
\label{fig:ZoR-3D-SPS}
\end{figure}

Then we apply the same evaluation procedure to determine the three-dimensional
reaction zone of the inelastic hadronic reactions ($2 \to 2' + m, m \geq 0$).
In Figs. \ref{fig:ZoR-3D-AGS}-\ref{fig:ZoR-3D-SPS}, the results of calculations
under conditions at the BNL Alternating Gradient Synchrotron (AGS), Au+Au, and at
the SPS, Pb+Pb, are depicted.
It is seen that, at SPS energies, the reaction zone (fireball) breaks-up into two
spatial parts.
Meanwhile, at AGS energies, there is virtually no break-up of a fireball,
but the fireball disappears almost immediately as a whole.
The reaction zones in the $z$-$r$ coordinates at times before and after the
break-up of a fireball are also shown in
Figs.~\ref{fig:ZoR-3D-AGS}-\ref{fig:ZoR-3D-SPS}.

While comparing the results of calculation for the energy $E_{\rm kin}=$158A~GeV
for all reactions (Fig. \ref{fig:Zor-trz}) and for inelastic reactions
(Fig. \ref{fig:ZoR-3D-SPS}), it is obviously seen that the reaction zone
containing 90\% of all reactions roughly coincides with the reaction zone
containing 99\% of all inelastic reactions.
We name this part of the reaction zone as a {\it hot fireball}.
Hence, everything that is beyond a hot fireball contains approximately 10\%
of the total number of reactions and just 1\% of all inelastic reactions.


In this study, we parameterize reaction zones by the global parameter $\alpha$
(fraction of the total number of reactions contained in the corresponding
reaction zone), rather than the local critical inelastic reaction
density $\Gamma_c$ for various collision energies.
We remind, that the hypersurface which separates different 4D zones is usually
determined, for instance, for the density of particles as $n(t,\bs r)=n_c$;
in our case this hypersurface can be determined as $\Gamma(t,\bs r)=\Gamma_c$.
The parameters $\alpha$ and $\Gamma_c$ are related to one another via
Eq.~\eqref{eq:gammac-alpha} for each particular collision energy.
It is interesting to explore this relation in greater details.
The values of $\Gamma_c$ for given different values of $\alpha$ and different
collision energies are presented in Table~\ref{tab:Gammac}.
It is seen that the values of $\Gamma_c (\alpha)$ are different at different
collision energies.

Hence, our approach with fixed $\alpha$ (which is a global parameter) for
different collision energies is not exactly equivalent to that with fixed
$\Gamma_c$ (which is a local quantity).
However, it is seen from Table~\ref{tab:Gammac} that the values of $\Gamma_c$
for the same given $\alpha$ are of the same order of magnitude for different
energies, whereas the change of $\alpha$ from 0.8 to 0.99 leads to a change
in $\Gamma_c$ of two orders of magnitude.
Therefore, the differences in $\Gamma_c$ for different energies are not very
significant.
This means that the approach, where $\Gamma_c$ is fixed, should yield
qualitatively the same results as those in the case where the fixed quantity
is $\alpha$ because the inverse dependence of $\alpha$ on $\Gamma_c$, i.e.
$\alpha(\Gamma_c)$, is weak.
We have performed calculations of the three-dimensional inelastic reaction zone
for different energies with $\Gamma_c (\alpha=0.99)$ values from
Table~\ref{tab:Gammac}, see Figs. \ref{fig:ZoR-3D-AGS}-\ref{fig:ZoR-3D-SPS}.
The resulting reaction zones for energies from Table~\ref{tab:Gammac} are
roughly the same in the range of these $\Gamma_c$ values, at least till the time
moment when the reaction zone separates into two different spatial pieces (drops).

\begin{table}
 \caption{{\bf Critical density $\Gamma_c$ of the inelastic reactions for given
               values of $\alpha$ }} %title
 \centering                                                 %centering table
 \begin{tabular}[c]{c|c|c|c|c}                              %columns {format}
  $E_{\rm kin}$& $\sqrt{s_{AA}}$ & $A+A$ &   $\Gamma_c~(\alpha=0.8)$ &
                                               $\Gamma_c~(\alpha=0.99)$\\
 (A~GeV) &  (GeV) &  &  $\left[\rm{fm}^{-3} \cdot (\rm{fm}/c)^{-1}\right]$ &
                            $\left[\rm{fm}^{-3} \cdot (\rm{fm}/c)^{-1}\right]$\\
 \hline\hline                                               %horizontal lines
 10.8  &  4.88 & $Au+Au$ & $119.07 \cdot 10^{-3}$ & $0.51 \cdot 10^{-3}$ \\
 \hline
 20.0 & 6.41   & $Pb+Pb$ & $83.63 \cdot 10^{-3}$  & $0.35 \cdot 10^{-3}$ \\
 40.0 & 8.86   &         & $60.59 \cdot 10^{-3}$  & $0.22 \cdot 10^{-3}$ \\
 80.0 & 12.39  &         & $42.86 \cdot 10^{-3}$  & $0.15 \cdot 10^{-3}$ \\
 158.0 & 17.32 &         & $31.42 \cdot 10^{-3}$  & $0.11 \cdot 10^{-3}$ \\
 \hline
 \end{tabular}
 \label{tab:Gammac}
\end{table}

Henceforth, we will deal with the projection of the reaction zone on the $z$-$t$
plane.
To design this projection, we sum firstly all collisions along the transverse
direction at the fixed coordinates ($t$, $z$).
Then, the reaction density in the $z$-$t$ plane takes the form
\begin{equation}
\widetilde{\Gamma} (t, z) = \int dx\, dy\, \Gamma(t, x, y, z)
= 2 \pi \int \, dr \, r \, \Gamma(t,r,z).
\label{Gamma-tz}
\end{equation}
Then the number of reactions in the given pixel
$\widetilde \Omega(t,z)$ on $z$-$t$ plane is
\begin{equation}\label{Ncoll-tz}
\widetilde N_{\rm coll}[\widetilde \Omega(t,z)]
= \int_{\widetilde \Omega(t,z)} dt\,dz \, \widetilde \Gamma(t,z) \,.
\end{equation}
To construct the reaction zone projection from UrQMD we divide the
two-dimensional $z-t$ plane into rectangular cells (pixels) with lengths
$\Delta t$=1~fm/$c$ and $\Delta z$=1~fm.
The volume of a pixel is then $\widetilde \Omega(t_i,z_j) = \Delta t \Delta z$
and does not depend on values of $t_i$ and $z_j$, i.e. it is the same for all
pixels.
It is different from the case of $(t,r,z)$ coordinates where the four-volume of
a pixel depended on transverse coordinate $r$.
This allows us to construct the reaction zone projection with the use of the
above-mentioned algorithm where in place of the pixel intensity $\Gamma(t,r,z)$,
which in the previous sorting problem determined the height of the bin, we put
the number of collisions in the correspondent pixel
$\widetilde N_{\rm coll}[\widetilde \Omega(t,z)]$, which is now the height of the
bin
(see details in Ref.~\cite{Anchishkin2010}).

It is worth to note that this reaction zone projection onto the $z$-$t$ plane
does not necessarily coincide with the corresponding projections of the
three-dimensional reaction zone in
Figs. \ref{fig:ZoR-3D-AGS}-\ref{fig:ZoR-3D-SPS}, which are just orthographic
projections of the reaction 4-density $\Gamma(t,r,z)$.
However, they are found to be roughly the same.

\begin{figure}[!t]
\begin{minipage}{.48\textwidth}
\centering
\includegraphics[width=\textwidth]{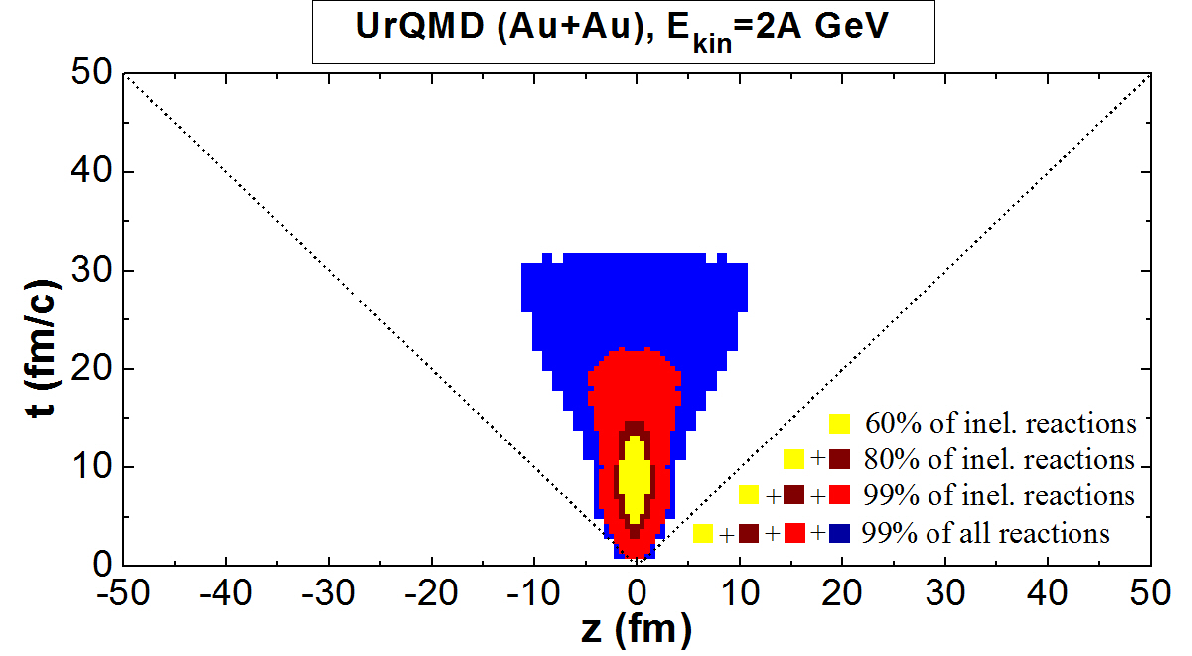}
\end{minipage}
 \rule{.02\textwidth}{0pt}
\begin{minipage}{.48\textwidth}
\centering
\includegraphics[width=\textwidth]{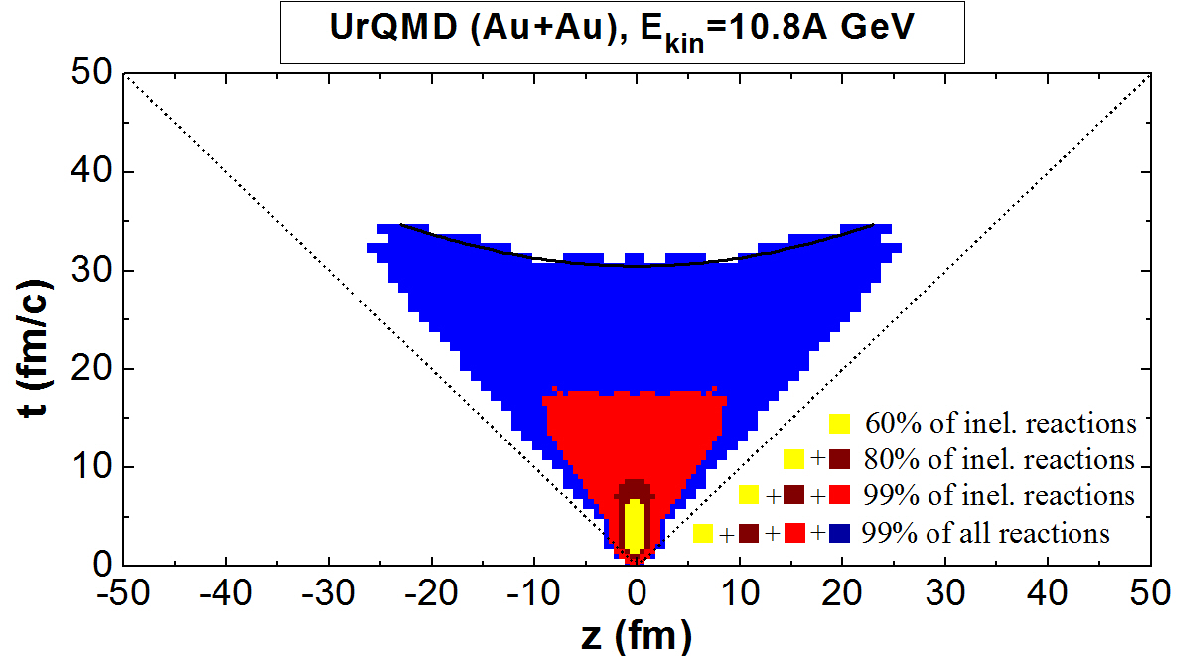}
\end{minipage}
\centering
\caption{Projection of the reaction zone on the $z$-$t$
plane under the AGS (Au+Au at $2A$~GeV and $10.8A$~GeV) conditions.
The yellow region contains 60\% of all inelastic reactions, $2\to 2'+m, m\ge 0$.
The dark red region together with previous region contains 80\% of all inelastic
reactions.
The red region together with the previous region contains 99\% of all inelastic
reactions.
The blue region together with the previous region contains 99\% of all hadronic
reactions.  }
\label{fig:ZoR-AGS}
\end{figure}

\begin{figure}[!t]
\begin{minipage}{.48\textwidth}
\centering
\includegraphics[width=\textwidth]{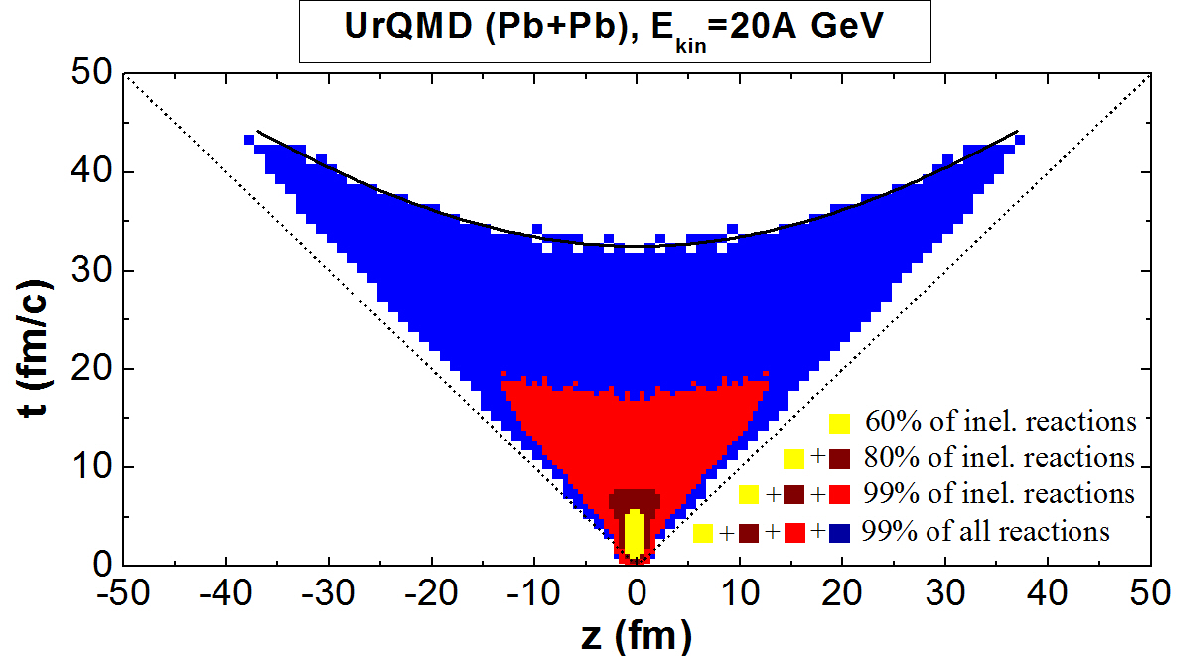}
\end{minipage}
 \rule{.02\textwidth}{0pt}
\begin{minipage}{.48\textwidth}
\centering
\includegraphics[width=\textwidth]{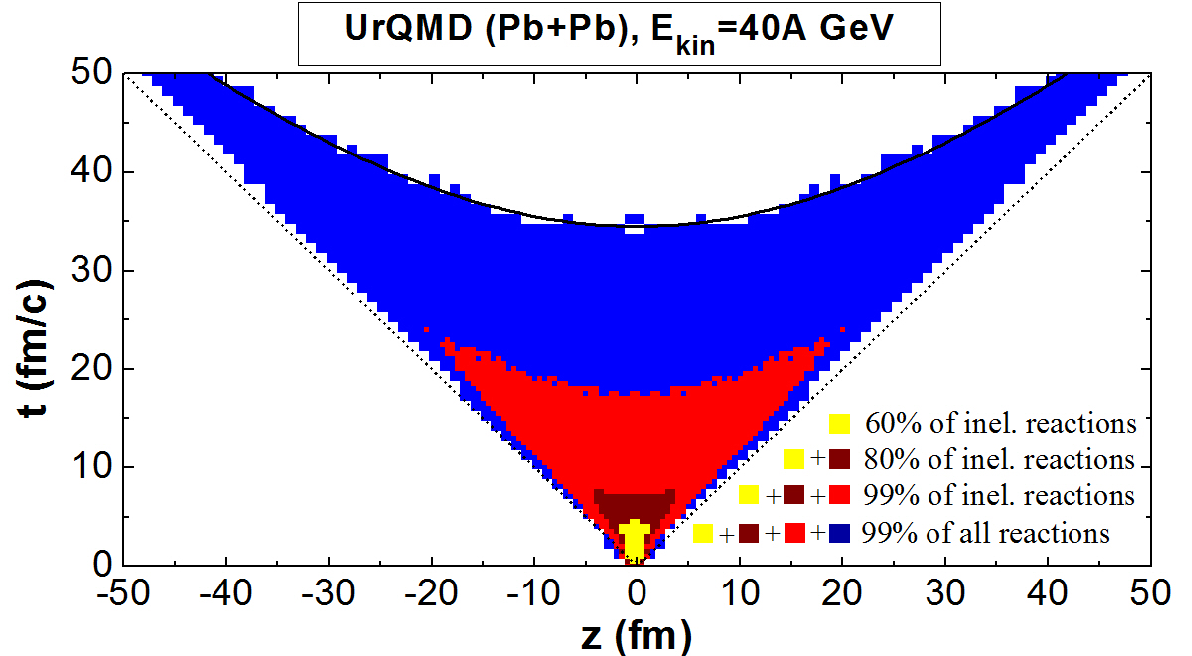}
\end{minipage}
\centering
\caption{Same as in Fig.~\ref{fig:ZoR-AGS},
   but for calculations under low SPS conditions (Pb+Pb at $20A$~GeV and $40A$~GeV).}
\label{fig:ZoR-lowSPS}
\end{figure}

\begin{figure}[!t]
\begin{minipage}{.48\textwidth}
\centering
\includegraphics[width=\textwidth]{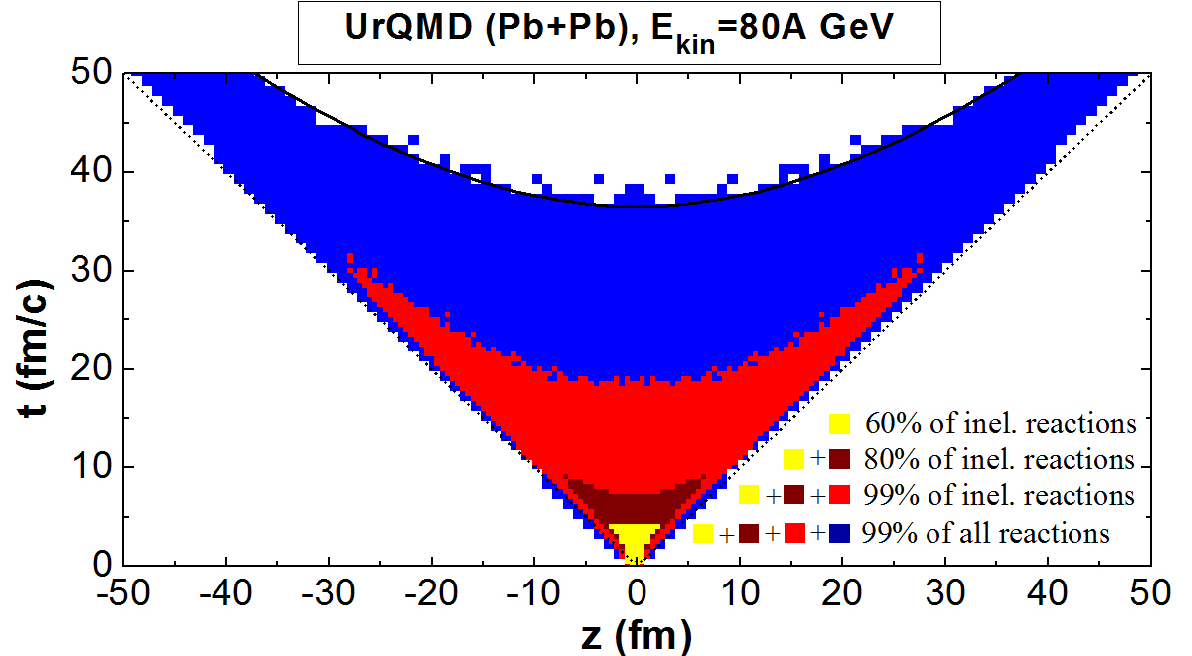}
\end{minipage}
\rule{.02\textwidth}{0pt}
\begin{minipage}{.48\textwidth}
\centering
\includegraphics[width=\textwidth]{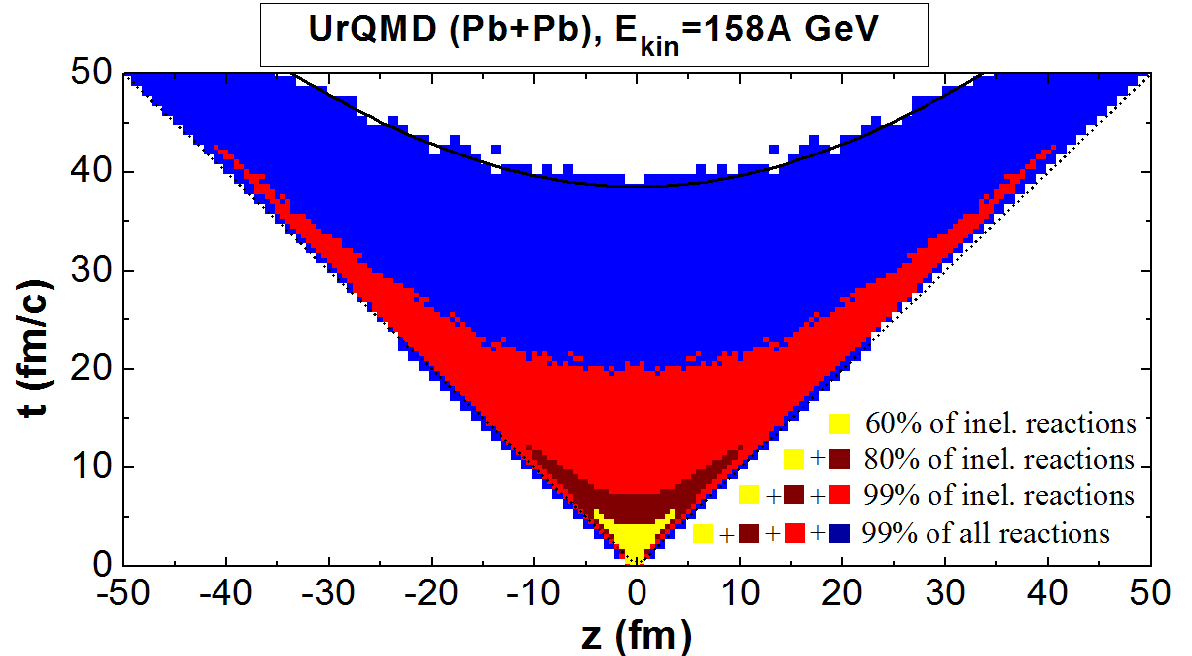}
\end{minipage}
\centering
\caption{Left and center -- same as in Fig.~\ref{fig:ZoR-AGS}, but for
         calculations under SPS conditions (Pb+Pb at $80A$~GeV and $158A$~GeV).}
\label{fig:ZoR-SPS}
\end{figure}

In Figs.~\ref{fig:ZoR-AGS}, \ref{fig:ZoR-lowSPS}, and \ref{fig:ZoR-SPS}, we show
the results of calculations under conditions at the BNL-AGS, Au+Au, and at the
CERN-SPS, Pb+Pb, in the case of central collisions.
In accordance with our algorithm, the volume that contains 60\% of all hadronic
inelastic reactions, $2\to 2' + m, m \ge 0,$ is determined (depicted as the
yellow area).
Next, we determine the volume that contains 80\% of all hadronic inelastic
reactions and includes the previous zone (depicted as the dark red area plus
the previous area).
Finally, we determine the volume that contains 99\% of all hadronic inelastic
reactions and also includes the previous zone (depicted as the red area plus the
previous two areas).
As was previously mentioned, this zone is named as the region of a hot fireball.
We determine also a volume that contains 99\% of all possible hadronic reactions
and includes, of course, the previous zone.
We name the region of 99\% of all hadronic reactions excluding the zone of the
hot fireball as a cold fireball (blue area).


\begin{figure}[!t]
\centering
\begin{minipage}{.35\textwidth}
\centering
\includegraphics[width=\textwidth]{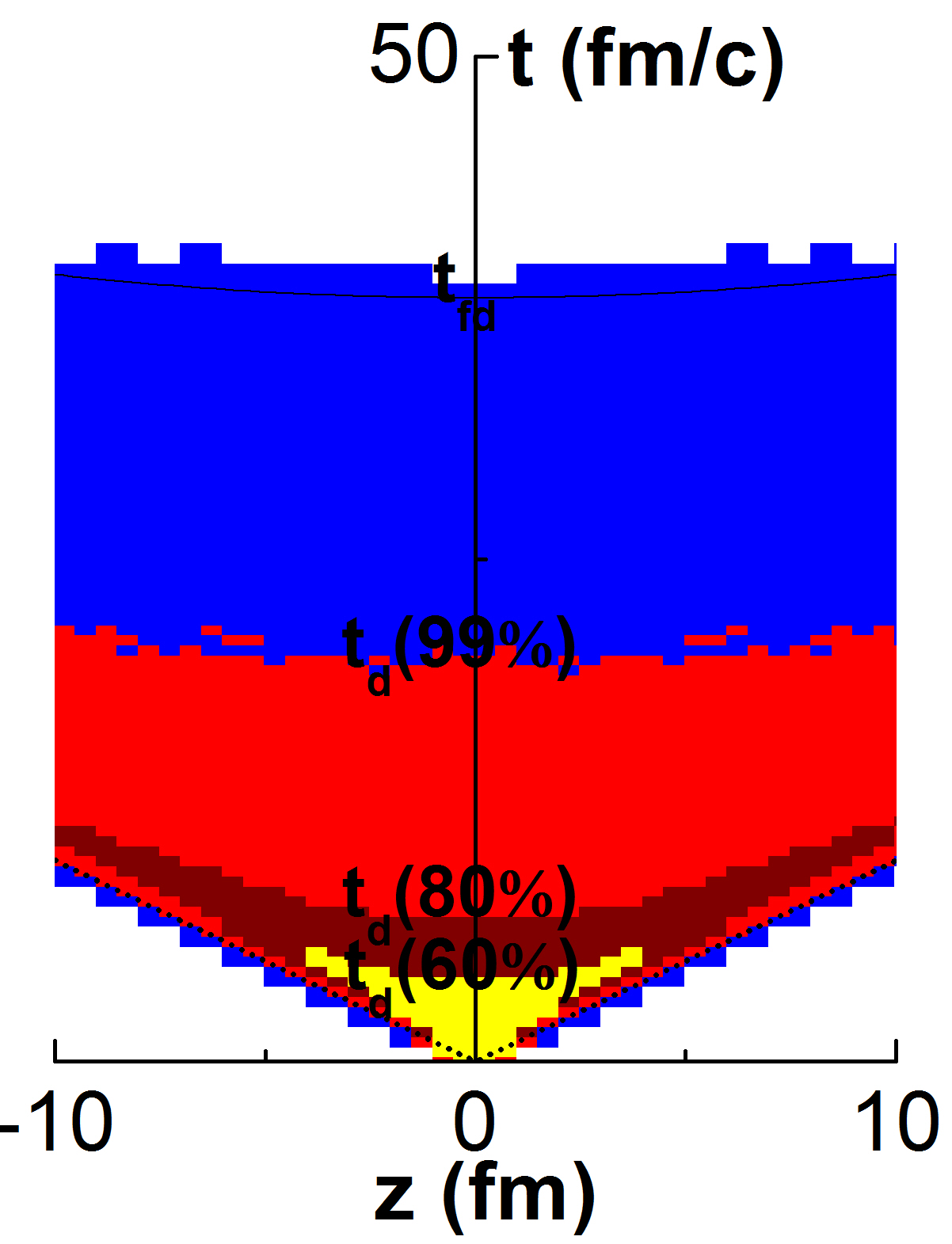}
\end{minipage}
\rule{.02\textwidth}{0pt}
%\centering
\begin{minipage}{.5\textwidth}
    \begin{center}
    \caption{Central part of the reaction zones under SPS conditions
    (Pb+Pb at $158A$~GeV),    where the inelastic reaction zone division times
    $t_{\rm d}$ are indicated for different values of $\alpha\times 100\%$:
    60\%, 80\%, and 99\%.}
    \label{fig:ZoR-SPS-2}
    \end{center}
    \end{minipage}
\end{figure}

After some time $t_{\rm d}$, the inelastic reaction zones studied in this work
either decay completely,
or are spatially divided into two parts, which move away from each
other in opposite directions, see Figs.~\ref{fig:ZoR-AGS}-\ref{fig:ZoR-SPS}.
This division time $t_{\rm d}$, which depends on the fraction $\alpha$,
can be defined as
$t_{\rm d} \equiv t_{\alpha}(z)\big|_{z=0}$, where the time dependence
$t_{\alpha}(z)$ defines the border between two reactions zones.
For details see Fig.~\ref{fig:ZoR-SPS-2} which represent in big scale the
corresponding reaction zone in the central region.
The values of $t_{\rm d}$ for different values of $\alpha$ and different
collision energies are depicted in Table~\ref{tab:trz}.

\begin{table}
 \caption{{\bf Division time $t_{\rm d}$ of the inelastic reaction zone for
               given values of $\alpha$  }} %title
 \centering                                                 %centering table
 \begin{tabular}[c]{c|c|c|c|c}                              %columns {format}
  $E_{\rm kin}$\!& $\sqrt{s_{AA}}$ %& $A+A$
  &   \!$t_{\rm d}(\alpha\!=\!0.6)$\! &   \!$t_{\rm d}(\alpha\!=\!0.8)$\! &
                                      \!$t_{\rm d}(\alpha\!=\!0.99)$\! \\
 \!(A~GeV)\! &  \!(GeV)\! %&
 &  (fm/$c$) & (fm/$c$) & (fm/$c$)\\
 \hline\hline                                               %horizontal lines
 10.8  &  4.88 %& $Au+Au$
 & 6.5 & 8.5 & 17.5\\
 \hline
 20.0 & 6.41   %& $Pb+Pb$
 & 5.5 & 7.5 & 17\\
 40.0 & 8.86   %&
 & 4.5 & 7 & 17.5\\
 80.0 & 12.39  %&
 & 4.25 & 7 & 18\\
 158.0 & 17.32 %&
 & 4.25 & 7 & 19.5\\
 \hline
 \end{tabular}
 \label{tab:trz}
\end{table}

One of the main features seen from Table~\ref{tab:trz} is that the division time
$t_{\rm d}$ of a separation of the particular reaction zone into two different
spatial parts depends weakly on the collision energy for different values
of $\alpha$.
That is especially true for higher values of $\alpha$ and in the case where the
reaction zones are divided into two parts at $t_{\rm d},$ rather than in the
case where they would decay completely
(see Figs.~\ref{fig:ZoR-AGS}-\ref{fig:ZoR-SPS}).

Similar result was also reported in studies of space-time structure of multipion
system created in heavy-ion collisions \cite{Anchishkin:2012jk}.
In that work the hypersurfaces which correspond to some invariant constant
densities were studied.
These hypersurfaces are defined by equations $n(t,\bs r) = n_c$ and
$\epsilon(t, \bs r) = \epsilon_c$, where $n_c$ is the critical number density
and $\epsilon_c$ the critical energy density of pi-mesons created in heavy-ion
collision.
It was found that for each particular value of critical density $n_c$
(or $\epsilon_c$), the division time $t_{\rm fd}$ depends weakly on
collision energy similarly to the case of reaction zones studied in this work.
These results indicate that {\it the weak dependence of fireball division time}
$t_{\rm fd}$ {\it on collision energy is a universal feature of space-time
structure of heavy-ion collisions}.

The zone including the hot and cold fireballs together contains 99\% of all
reactions by definition (see the colored area in
Figs.~\ref{fig:ZoR-AGS}-\ref{fig:ZoR-SPS}).
We can compare this reaction zone with the zone which is wrapped by
the freeze-out hypersurface.
Following the ``classical'' definition of the sharp kinetic
freeze-out hypersurface, it is some boundary that separates the
interacting system from the space-time domain, where the particles do not
interact, and almost all particles are evaporated (freezed out) from
the thin space-time layer determined by this hypersurface.
Then, it seems evident that the sharp kinetic freeze-out hypersurface
should be inside the reaction zone, rather than outside it.
In addition, we note that the sharp chemical freeze-out hypersurface should be
inside the hypersurface that separates the region of a hot fireball
containing 99\% of all inelastic reactions which corresponds to the area
covered by the first three zones together depicted in
Figs.~\ref{fig:ZoR-AGS}-\ref{fig:ZoR-SPS}.

In the coordinates $(t,\,z)$, the curve which is the upper space-like
boundary of the cold fireball (``blue'' zone) can be well approximated as a
$\tau_{_{\rm RZ}} = {\rm const}$ hyperbola originating from a possibly
different time, $t^0_{_{\rm RZ}}$, than the initial time $t=0$ of the
collision. This approximation was already used to describe the $[t,\, z]$-projection of
hypersurface of constant invariant pion density~\cite{Anchishkin:2012jk}.
The equation for this hyperbola has the form
\begin{equation}
t_{_{\rm RZ}}(z)\ =\ t^0_{_{\rm RZ}}\, +\,
\sqrt{\tau_{_{\rm RZ}}^2\, +\, z^2} \,.
\label{eq:tz-appro}
\end{equation}
The fireball division time $t_{\rm fd}$ is related to hypersurface parameters as
\begin{equation}
t_{\rm fd} = t^0_{_{\rm RZ}} + \tau_{_{\rm RZ}}.
%\label{eq:tz-appro}
\end{equation}
The values of parameters $\tau_{_{\rm RZ}}$ and $t^0_{_{\rm RZ}}$ for different energies
are presented in Table~\ref{tab:hypersparams2}. It is seen that the parameter $t^0_{_{\rm RZ}}$ takes negative
values and approaches zero with increase of the collision energy. In that case,
the reaction zone boundary can be regarded
as the hypersurface of a constant proper time $\tau_{_{\rm RZ}}$, which originates from
initial time $t=0$ and which is then equal to $t_{\rm fd}$.
It should be mentioned that coordinate $z$ in Eq. \eqref{eq:tz-appro}
takes values only in finite
interval $-z_{\rm max}<z<z_{\rm max}$, where $\pm z_{\rm max}$ is $z$-coordinate
of the intersection point between upper space-like boundary of the reaction zone
and lower time-like boundary.

The upper space-like boundary of the ``blue'' zone can
be parameterized in a different way. For instance, it
can be approximated as
a hyperbola of the form $t(z)=A\sqrt{{\tau_0}^2+z^2}$ \cite{Anchishkin2010}, where
$A = 0.65$, ${\tau_0} = 46$~fm$/c$
at the AGS energy ($E_{\rm kin} = 10.8A$~GeV) and
$A = 0.8-0.95$, ${\tau_0} = 38-42.5$~fm$/c$
at the SPS energies ($E_{\rm kin} = 20-158A$~GeV).
Here, parameter $A$
approaches unity with increase of the collision energy,
and the space-like boundary takes the form of the hypersurface of a constant proper time $\tau_{0}$, which originates from
initial time $t=0$ and coincides with previous parametrization at high energies.
At AGS and SPS energies, however, the space-like boundary has a more complex structure and
both parameterizations can be used to describe it.
We note, that the presented parameterizations can be used
to describe upper space-like boundaries of also other reaction zones, rather than ``blue'' zone alone,
provided that there is a spatial division of that zones (see different zones in Figs.~\ref{fig:ZoR-AGS}-\ref{fig:ZoR-SPS}).

\begin{table}
 \caption{{\bf Parameters of $\tau$-const. hyperbola \newline (space-like boundary of cold fireball) }}       %title
 \centering                                                             %centering table
 \begin{tabular}[c]{c|c|c}                                              %columns {format}
 \hline\hline                                                           %horizontal lines
 $E_{\rm kin}$ (A~GeV) & $t^0_{_{\rm RZ}}$ (fm/$c$) & $\tau_{_{\rm RZ}}$ (fm/$c$)\\
 \hline
 10.8 (AGS) & -30                  & 60.5 \\
 \hline
 20 (SPS) & -20                     & 52.5 \\
 \hline
 40 (SPS) & -14                    & 48.5 \\
 \hline
 80 (SPS) & -8                    & 44.5 \\
 \hline
 158 (SPS) & -5                   & 43.5 \\
 \hline
 \end{tabular}
 \label{tab:hypersparams2}
\end{table}

The time-like hypersurface bounding cold fireball from below has the
form of a straight line $t(z)=t_0+{1\over v}z$, where $t_0$ is
close to zero, and $v = 0.8$ at AGS energies and $v = 0.88-0.98$ at
SPS energies ($v$ increases with the collision energy).
At AGS energies, the time-like boundaries
of the reaction zones differ significantly from one another and the
light cone (see Fig. \ref{fig:ZoR-AGS}).
However, at higher SPS energies (for example, at
$E_{\rm kin} = 158A$~GeV), the time-like hypersurfaces bounding all
three zones of a fireball practically coincide with one another and
are close to the light cone (see Fig.~\ref{fig:ZoR-SPS}).
Thus, we can predict that
time-like hypersurfaces which bound different reaction zones from below
merge at the energies available at the
Relativistic Heavy Ion Collider or Large Hadron Collider and virtually coincide with
the light cone.

%%%%%%%%%%%%%%%%%%%%%%%%%%%%%%%%%%%%%%%%%%%%%%%%%%%%%%%%%%%%%%%%%%%%%%%%%%%%%%%%

%%%%%%%%%%%%%%%%%%%%%%%%%%%%%%%%%%%%%%%%%
\section{Temporal structure of reactions}
%%%%%%%%%%%%%%%%%%%%%%%%%%%%%%%%%%%%%%%%%
The structure of hadronic reactions in a fireball can be analyzed within
UrQMD, which allows one to calculate the reaction density for various types of
reactions.

The reactions can be classified by the type and the number of participants in a
reaction (see Table \ref{tab:reacttype}), and the contributions of different
types of reactions to the system evolution can be explored.
For this purpose, we analyze the time dependence of the reaction frequency for
different types, $i$, of reactions:
\begin{equation}\label{eq:freq}
\nu_i(t) = \int_{C_R} dx dy dz \, \Gamma_i(t, \bs r).
\end{equation}

\begin{figure}[!t]
\begin{minipage}{.48\textwidth}
\centering
\includegraphics[width=\textwidth]{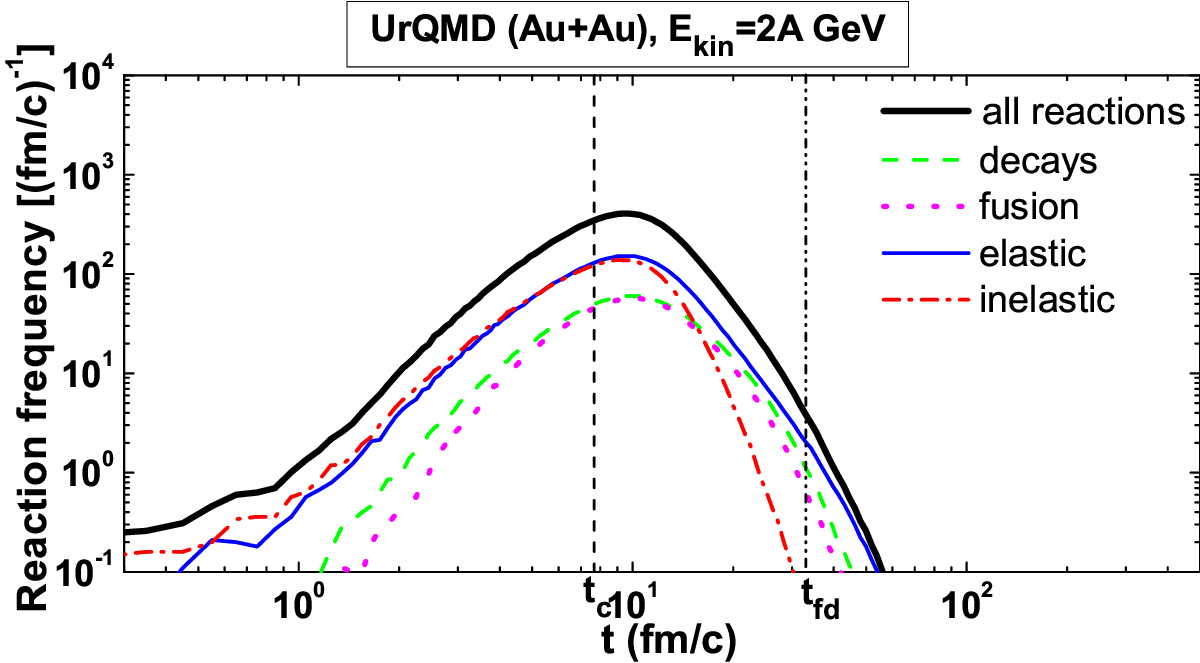}
\end{minipage}
 \rule{.02\textwidth}{0pt}
\begin{minipage}{.48\textwidth}
\centering
\includegraphics[width=\textwidth]{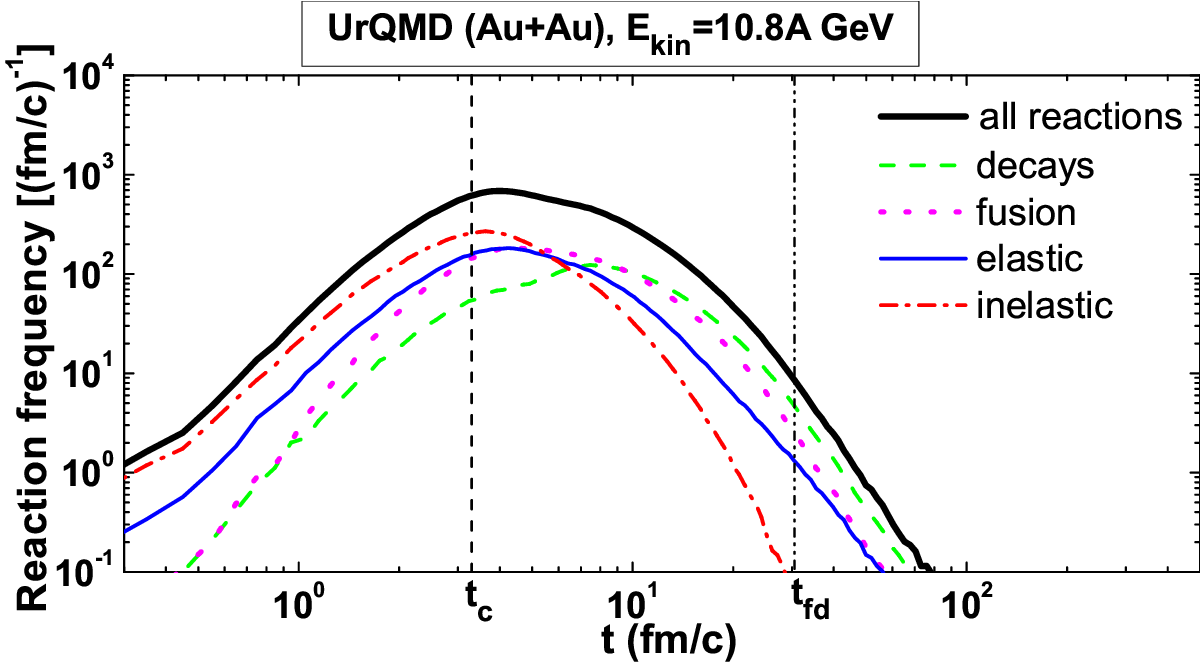}
\end{minipage}
\centering
\caption{Hadron reaction frequency under AGS conditions (Au+Au at $2$ and
$10.8A$~GeV).
Different curves correspond to different types of reactions.}
\label{fig:FoR-AGS}
\end{figure}

\begin{figure}[!t]
\begin{minipage}{.48\textwidth}
\centering
\includegraphics[width=\textwidth]{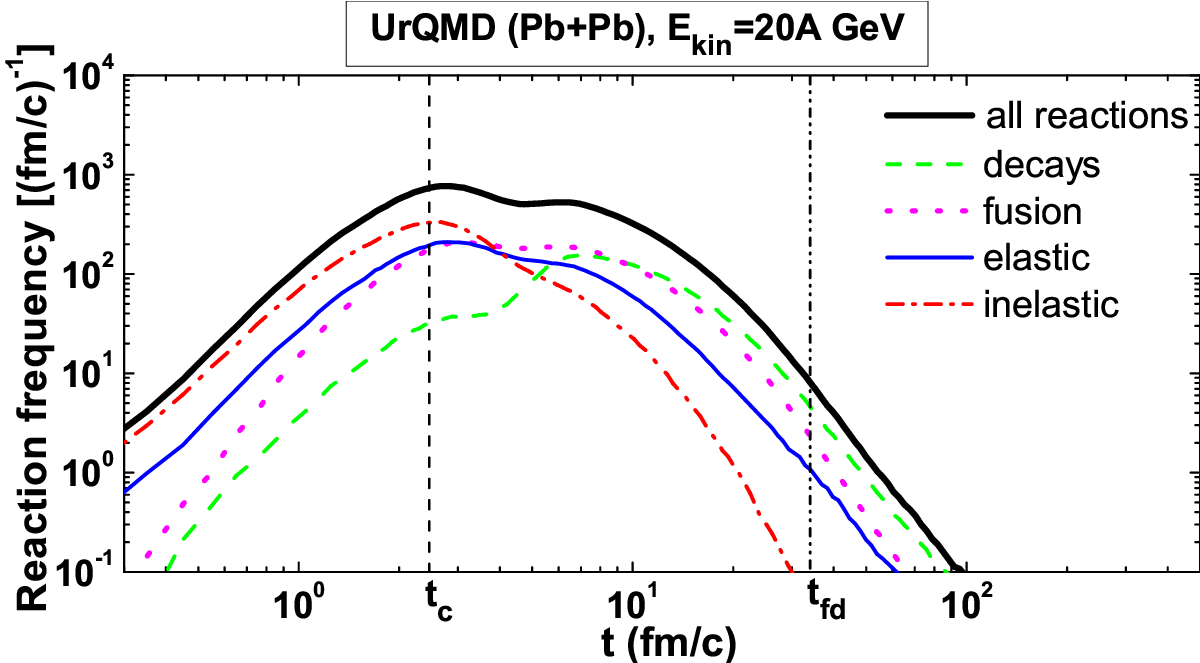}
\end{minipage}
 \rule{.02\textwidth}{0pt}
\begin{minipage}{.48\textwidth}
\centering
\includegraphics[width=\textwidth]{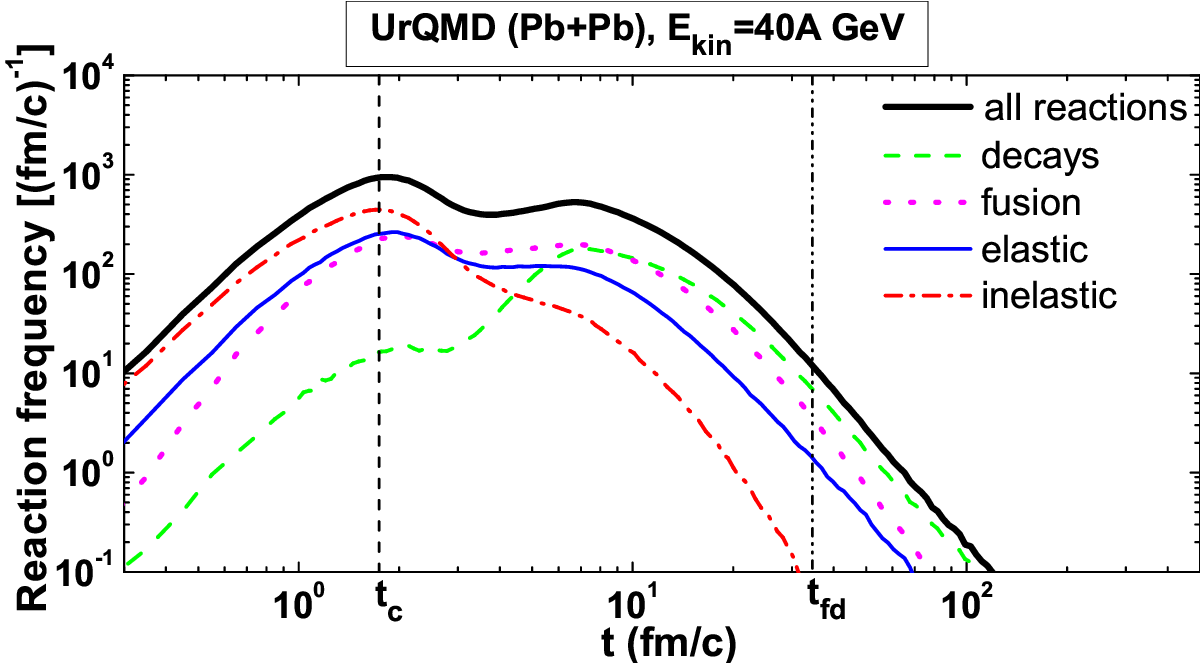}
\end{minipage}
\centering
\caption{Same as in Fig.~\ref{fig:FoR-AGS},
   but for calculations under low SPS conditions (Pb+Pb at $20A$~GeV and $40A$~GeV).}
   \label{fig:FoR-lowSPS}
\end{figure}

\begin{figure}[!t]
\begin{minipage}{.48\textwidth}
\centering
\includegraphics[width=\textwidth]{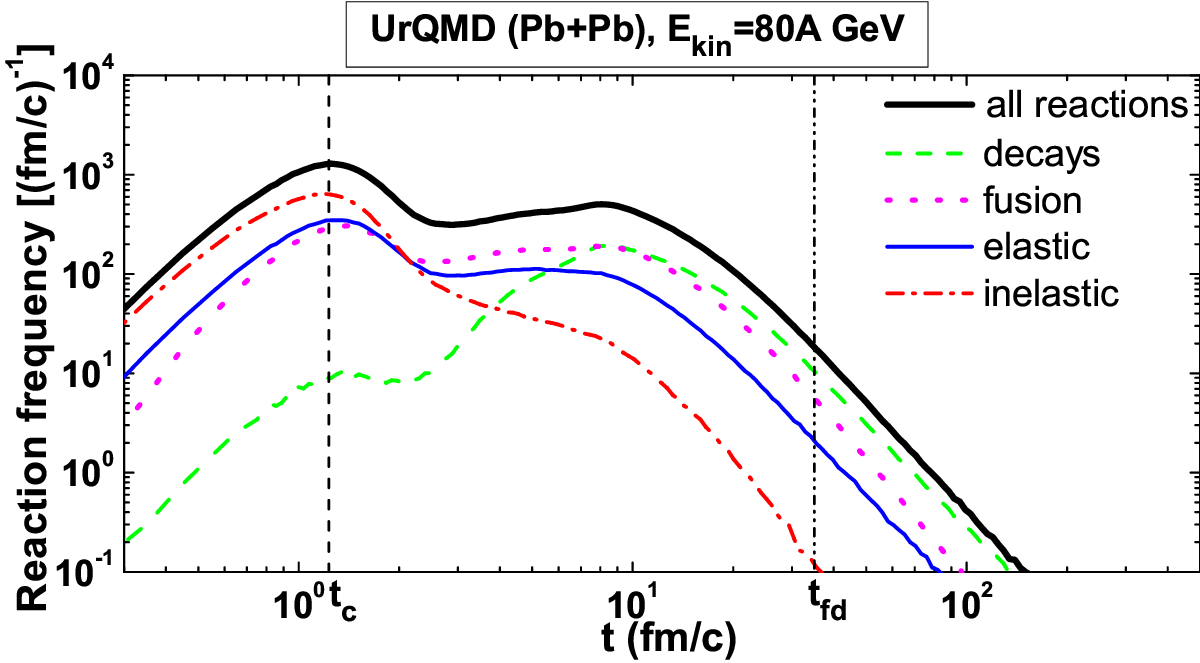}
\end{minipage}
 \rule{.02\textwidth}{0pt}
\begin{minipage}{.48\textwidth}
\centering
\includegraphics[width=\textwidth]{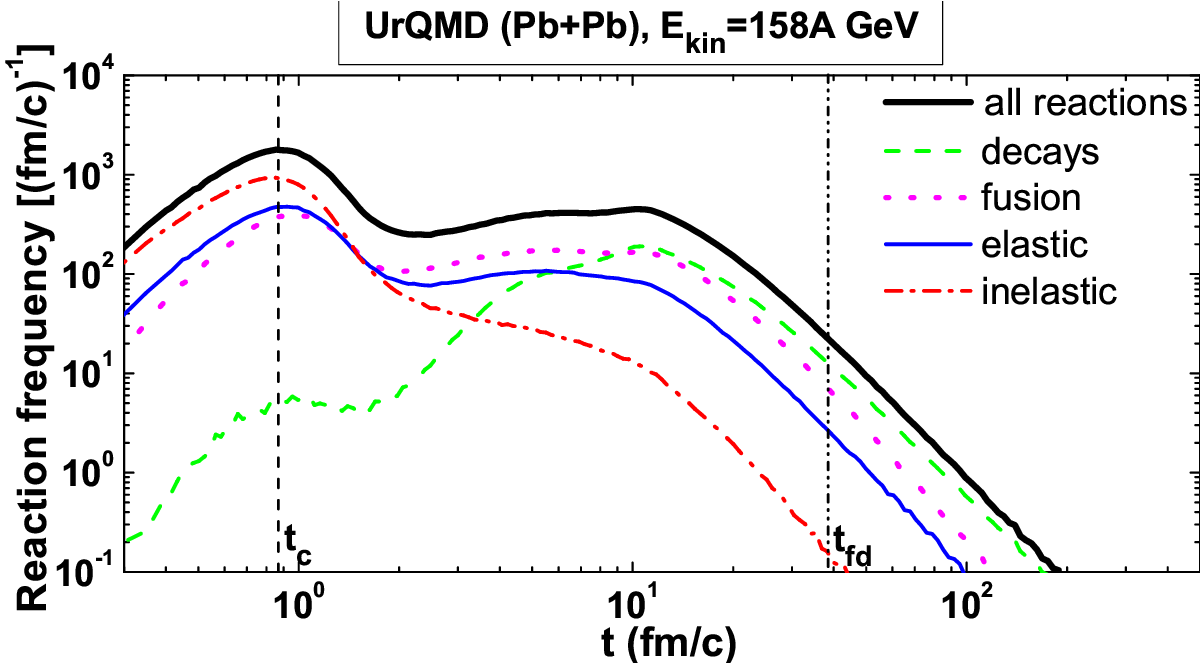}
\end{minipage}
\centering
\caption{Same as in Fig.~\ref{fig:FoR-AGS},
   but for calculations under SPS conditions (Pb+Pb at $80A$~GeV and $158A$~GeV).}
   \label{fig:FoR-SPS}
\end{figure}

The results of evaluations of the time dependence of the reaction frequency
(\ref{eq:freq}) at the AGS and SPS energies are depicted
in Figs.~\ref{fig:FoR-AGS}, \ref{fig:FoR-lowSPS} and \ref{fig:FoR-SPS}.
The thick solid line indicates all reaction rates in a fireball, the
thin solid line indicates only the elastic scattering of hadrons
($2\to 2$), the dash-dotted line shows all inelastic reactions ($2
\to 2'+m$, where $m \ge 0$), the dotted line stands for fusion
reactions ($2 \to 1'$), and the dashed line distinguishes decays ($1
\to 2'+m, m \ge 0$).

\begin{table}[!b]
 \caption{{\bf Temporal characteristics of reaction frequency.}} %title
 \centering                                                 %centering table
 \begin{tabular}[c]{c|c|c|c|c|c}                              %columns {format}
  $E_{\rm kin}$ (A GeV)& $\sqrt{s_{AA}}$ (GeV)%& $A+A$
  &$t_c$
  (fm$/c$)& $t_{m1}$ (fm$/c$)& $t_{m2}$ (fm$/c$)& $t_{fd}$ (fm$/c$)\\
 \hline\hline                                               %horizontal lines
 2.0   &  2.70 &  7.66  & 9.7  &   & 33.0 \\
 10.8  &  4.88 &  3.30  & 3.9  &    & 30.5  \\
 \hline
 20.0 & 6.41   & 2.46 & 2.73 & 6.1  & 34.0  \\
 40.0 & 8.86   & 1.74 & 1.85 & 6.7  & 34.5  \\
 80.0 & 12.39  & 1.23 & 1.27 & 8.2  & 35.0  \\
 158.0 & 17.32 & 0.87 & 0.88 & 10.3 & 38.5 \\
 \hline
 \end{tabular}
 \label{tab:FoR-time-points}
\end{table}

The main feature of the reaction frequency (thick solid lines in
Figs.~\ref{fig:FoR-AGS}-\ref{fig:FoR-SPS}) is its increase up to
$t \simeq 3.9-8$~fm$/c$ at AGS energies and $t \simeq 0.84-2$~fm$/c$ at SPS
energies where it has its first maximum $t_{\rm m1}$.
This can be explained by an increase of the number of nucleons as participants
of the reactions, when one nucleus penetrates into another one.
Indeed, the maximum overlap of two nuclei happens, when their centers coincide.
This time can be estimated as
\begin{equation}\label{tc}
t_c = \frac{R_{0}}{\gamma}  \frac 1v \,,
\end{equation}
where $R_0$ is the nucleus radius, $v = p_{0z}/\sqrt{M_N^2 + p_{0z}^2}$,
$\gamma = 1/{\sqrt{1 - v^2}}$, $p_{0z}$ is the initial nucleon momentum in the
center-of-mass system of two nuclei, and $M_N$ is the nucleon mass.
We call $t_c$ as the fireball formation time.
The values of $t_c$ (see Table \ref{tab:FoR-time-points}) are very close to the
time moments that correspond to the first maximum of the reaction frequency
(thick solid lines in Figs.~\ref{fig:FoR-AGS}-\ref{fig:FoR-SPS}).
A slight difference of $t_c$ and the time point of the real maximum
can be explained by some decrease of the nucleon velocity, which
is due to inelastic and elastic reactions (stopping) of nucleons.


As it is seen in Figs.~\ref{fig:FoR-AGS}-\ref{fig:FoR-SPS}, for all energies,
the inelastic and elastic nucleon collisions dominate at the first stage,
$t < t_c\,$, of a nucleus-nucleus collision.
At AGS energies, these reactions are
most dominant at all times except for very late time moments.
After the full overlap of the nuclei, the created system begins to expand in
space, which results in a decrease of the reaction frequency,
as the nucleon reactions and the reactions involving secondary
particles become less and less intense.

The structure of the reaction frequency changes at higher energies available
at SPS.
As the collision energy increases, the maximum overlap of colliding nuclei
happens at earlier times~(see Eq.~\eqref{tc} and
Figs.~\ref{fig:FoR-lowSPS}-\ref{fig:FoR-SPS}) due to a bigger initial velocity
and the Lorentz contraction of colliding nuclei
and more secondary particles are created at this stage.
The total contribution of the secondary particles becomes more significant at
SPS energies, than at AGS energies, due to an increase of the number of secondary particles with the
collision energy.
The number of secondary particles (mainly $\pi$ mesons) is approximately
$\langle n_\pi \rangle \simeq 0.6-1.6$ per nucleon under AGS conditions
\cite{Klay2003} and $\langle n_\pi \rangle \simeq 2-6$ under SPS
conditions \cite{Alt2008,Afanasiev2002}.
Frequency of hadronic reactions at subsequent times after nuclei overlap
may be somewhat specific to the implementation of string dynamics in UrQMD.
New hadronic particles are created in UrQMD through string fragmentation and
have a formation time, whereas the
dynamics of the partons during that formation time are not implemented in UrQMD.
The contribution of the unformed hadrons, where explicit treatment of partonic
degrees of freedom is important, to the energy density and collision dynamics of system
was studied within UrQMD model in Ref. \cite{Weber1998}. Those studies indicate
that, after the maximum overlap of nuclei, the partonic degrees become dominant
for some duration, hence this indicates the emergence of a partonic stage
and possible reason why the reaction frequency goes down at SPS energies
after nuclei overlap~(see Figs.~\ref{fig:FoR-lowSPS}-\ref{fig:FoR-SPS}).
The processes involving already formed
secondary hadrons (mainly, decays and fusions) are most intense at later times
in the interval of $6-10$~fm/$c$. Evidently, this causes the appearance of the second local maximum
of the hadronic reaction frequency at SPS energies (see $t_{m2}$ values in Table \ref{tab:FoR-time-points}), which coincides
with the local maxima of the decay and fusion reactions (see the thick solid line in
Figs.~\ref{fig:FoR-lowSPS} and \ref{fig:FoR-SPS}).
Similar behavior of the reaction frequency can be seen as well in Ref.~\cite{Bleicher2002},
Fig.~1, for Pb-Pb collisions at
$E_{\rm kin} = 158A$~GeV.

Naturally, we come to the prediction concerning higher energies.
With increase of the collision energy, the appearance of two local maxima indicates
a separation of the total frequency of hadron reactions into two parts:
the first part is mainly attributed to nucleon processes with the most
intensive reactions occurring when two colliding nuclei fully overlap, see Eq.~\eqref{tc},
and the second part is mainly attributed to decay, fusion, and elastic
reactions involving already formed secondary particles,
the total duration of the second stage gradually increases with the collision energy.
For accurate treatment of the intermediate stage, which emerges after the nuclei overlap,
an explicit account for partonic degrees of freedom appears to be necessary,
especially at higher collision energies.

At later times after $t=t_{\rm m2}$, the reaction frequency goes down, which
results in the division of a fireball into two spatial
parts at the time moment $t = t_{\rm fd}$ and in the further breakup.
The fireball division time is defined as the minimum value of time on the
space-like hypersurface, which bounds the region of the cold fireball
(blue area) from above, i.e., $t_{\rm fd}\equiv t(z)\big|_{z=0}$
(see Figs. {\ref{fig:ZoR-AGS}}-{\ref{fig:ZoR-SPS}}).
We note that the time moment $t_{\rm fd}$ depends weakly on the
collision energy (see Table \ref{tab:FoR-time-points}).
It is seen that, after the time moment
$t_{\rm fd}$, the rates of elastic and inelastic reactions vanish.
In other words, the system behavior is determined since this moment mainly
by the individual properties of particles (basically, by resonances).
That is why, in spite of the sufficient difference of collision
energies of the experiments under consideration, the times $t_{\rm
fd}$ are approximately the same (see Table
\ref{tab:FoR-time-points}).
For this reason, the longitudinal sizes of the fireballs $2 R_z$ at the time
moment of division into two separate parts, $t=t_{\rm fd}$, are approximately
the same and equal to $R_z \approx v \, t_{\rm fd}$ [$v$ is defined in
Eq.~(\ref{tc}), see Figs.~\ref{fig:ZoR-AGS}-\ref{fig:ZoR-SPS}].
This fact can explain the weak dependence of the pion interferometric radius
$R_{L}$ on the beam energy \cite{Adler2001STAR,Adams2005STAR} because
$R_{L} \propto R_z$.
So, it can be claimed that the fireball achieves its maximum longitudinal size as one spatial object
at the time moment $t=t_{\rm fd}$, when it is divided into two different spatial
parts.

%%%%%%%%%%%%%%%%%%%%%%%%%%%%%%%%%%%%
\section{Discussion and conclusions}
%%%%%%%%%%%%%%%%%%%%%%%%%%%%%%%%%%%%

Different parameters such as the energy density, particle density, mean free
path, etc., can be used to analyze the fireball evolution.
Our approach allows one to investigate the spatial and temporal structures
of the hadron system created in relativistic nucleus-nucleus collisions
in terms of hadronic reactions that occur in the system.
In other words, the fireball is identified as a system of interacting hadrons.
The proposed algorithm gives possibility to separate, with a given
accuracy, the space-time region, where the most intense hadron reactions
take place, i.e. we give the method to see a reaction zone in the 3D
representation
(see Fig.~\ref{fig:Zor-trz}) and in different projections.

In the present microscopic study, we separate a fireball into the following
regions, which characterize its evolution (see
Figs.~\ref{fig:ZoR-AGS}-\ref{fig:ZoR-SPS}):
(1) a fireball region, where 60\% of all inelastic hadronic
reactions have occurred (yellow area),
(2) a fireball region, where 80\% of all inelastic hadronic
reactions have occurred (yellow plus dark-red area),
(3) a hot fireball region, where 99\% of all inelastic hadronic
reactions have occurred (yellow plus dark-red plus red area),
(4) a cold fireball region (blue area), which together
with the hot fireball contains 99\% of all hadronic reactions $N_{\rm tot}$.
The last region  (blue area in Figs.~{\ref{fig:ZoR-AGS}}-{\ref{fig:ZoR-SPS}})
contains the hadron-resonance gas, and the reactions in this region are mainly
presented by decays of resonances if we consider times $t \ge t_{\rm fd}$, see
Figs.~{\ref{fig:FoR-AGS}}-{\ref{fig:FoR-SPS}}.

The study of hadron reaction zones allows one to analyze the freeze-out
process in relativistic nucleus-nucleus collisions.
Indeed, in the literature, the sharp freeze-out hypersurface is usually defined
with the
help of some parameter $P(t,\bs r),$ which takes the critical value $P_c$ on the
hypersurface, i.e. the equation for that hypersurface reads as $P(t,\bs r)=P_c$.
For instance, this quantity can be chosen as particle density $n(t,\bs r)$
\cite{CERES,Anchishkin:2012jk}, energy density $\epsilon(t,\bs r)$
\cite{Anchishkin:2012jk,RusskikhIvanov2006}, temperature $T(t,\bs r)$
\cite{mclerran1986,Huovinen2007}, etc.
Moreover, the ``classical'' definition of sharp kinetic freeze-out assumes the
Cooper--Frye picture \cite{cooper-frye-PRD-v10-1974}:
a radiation of free particles or the freeze out process takes place within a
thin layer determined by a freeze-out hypersurface;
this approach is usually used to describe transition from the fluid dynamical
stage of heavy-ion collision to the stage of dilute hadron gas.
That is why, the initial value problem for the radiation of free hadrons is
formulated with the use of the space-like piece of the hypersurface, whereas the
boundary conditions are formulated exploiting the time-like part of the
hypersurface.
Thus, the knowledge of the freeze-out hypersurface is important for estimation
of the spectrum of secondary particles, for instance pions.
On the other hand, if we determine the reaction zone as that containing, for
example, 99\% of all reactions, then we can claim that the sharp kinetic
freeze-out hypersurface should be definitely inside this zone (see
Figs.~\ref{fig:ZoR-AGS}-\ref{fig:ZoR-SPS})
while the parametrization
of iso-$\tau$ hypersurface of the form given in Eq. \eqref{eq:tz-appro} can be used to define this hypersurface.
Indeed, as was found in \cite{Anchishkin:2012jk} the pion freeze-out
hypersurface which is calculated from equation $n_\pi(t,\bs r)=n_c$, where
$n_c=0.08$~fm$^{-3}$, approximately coincides with the hypersurface which bounds
reaction zone that contains 80\% of all inelastic hadronic reactions.
(Here the value of $n_c$ is chosen from the condition that the
freeze-out size of the pion system is to be adequate to known HBT radius.)
Assuming that the chemical freeze-out occurs, when the inelastic reactions are
completed (see Ref.~\cite{Heinz2001}), we can also claim that the chemical
freeze-out hypersurface should be inside the reaction zone, which contains
99\% of all inelastic reactions.

While the Cooper-Frye picture is convenient to use and has many
applications in heavy-ion collisions, it is also a rather
crude idealization of the freeze-out process.
The more realistic approach could be a description of the freeze-out process from the extended space-time
layer (volume) rather than from the space-time hypersurface \cite{Molnar2006,Molnar2007}.
Considering different reaction zones in particular projections, which are presented in our work,
we can conclude that such a freeze-out layer is most extended in time in the central region
and, especially at higher collision energies, it becomes very narrow at the sides of $t$-$z$ projection, where the amount of matter which can
leave the interacting system becomes minimal~(see different areas in Figs.~\ref{fig:ZoR-AGS}-\ref{fig:ZoR-SPS}).

By studying the time dependence of the reaction frequency for
different reaction types, we conclude that the total hadronic reaction rate
is dominated by elastic and inelastic hadron collisions at the early
stage, whereas the individual properties of particles (basically,
resonances) and reactions involving secondary particles determine
the behavior of the system at later stages.
The initial stage is characterized by the fireball formation time $t_c$ which is
defined as the time of the full overlap of two nuclei and thus it is
determined by collision energy [Eq.~(\ref{tc})].
With increase of collision energy this time moment is a good estimation of the
time moment $t_{m1}$, which is a maximum of nucleon-nucleon reaction frequency.
During the initial stage a hot and dense nuclear matter is created,
while the partonic degrees of freedom play more important role in subsequent evolution
of the system, especially at higher collision energies.
At later times, the hadronic evolution is determined mainly by decay and fusion reactions which are
most intense at time moment $t_{m2}$ (see Table~\ref{tab:FoR-time-points}), and
the total duration of this stage increases with collision energy.
Hence, the features of the second stage are mainly determined by the individual
properties of secondary particles.
Another specific time point in the evolution of a hadron fireball is the
fireball division time $t_{\rm fd}$, which corresponds to the separation of
the fireball into two different spatial parts and depends weakly on the
collision energy (see the last column in Table~\ref{tab:trz}).
%

%%%%%%%%%%%%%%%%%%%%%%%%%%%
\section*{Acknowledgements}
%%%%%%%%%%%%%%%%%%%%%%%%%%%
\noindent
Authors are thankful to L. Csernai for useful discussions
and to L. Bravina for valuable comments.
D.A. was supported by the program \textquotedblleft
Microscopical and phenomenological models of fundamental physical processes at
micro and macro scales\textquotedblright\
(Section of physics and astronomy of the NAS of Ukraine).

%%%%%%%%%%%%%%%%%%%%%%%%%

\end{document}